\documentclass[reprint,aps,prb,superscriptaddress,nofootinbib,longbibliography]{revtex4-1}

\usepackage[utf8]{inputenc}
\usepackage{verbatim}
\usepackage{color}
\usepackage{times}
\usepackage{amsmath}
\usepackage{amssymb}
\usepackage{stmaryrd}
\usepackage{makecell}
\usepackage[caption=false]{subfig}
\usepackage{xr-hyper}
\usepackage{graphicx}
\usepackage{tikz-cd}
\usepackage{adjustbox}
\usepackage[unicode=true,
 bookmarks=true,bookmarksnumbered=false,bookmarksopen=false,
 breaklinks=false,pdfborder={0 0 1},backref=false,colorlinks=true]
 {hyperref}
\hypersetup{
 linkcolor=magenta, urlcolor=blue, citecolor=blue, pdfstartview={FitH}, hyperfootnotes=true, unicode=true}
\makeatletter

\@ifundefined{textcolor}{}
{%
 \definecolor{BLACK}{gray}{0}
 \definecolor{WHITE}{gray}{1}
 \definecolor{RED}{rgb}{1,0,0}
 \definecolor{GREEN}{rgb}{0,1,0}
 \definecolor{BLUE}{rgb}{0,0,1}
 \definecolor{CYAN}{cmyk}{1,0,0,0}
 \definecolor{MAGENTA}{cmyk}{0,1,0,0}
 \definecolor{YELLOW}{cmyk}{0,0,1,0}
 \definecolor{dkgray}{gray}{0.4}
}

\newcommand{\tr}{\mathop{\mathrm{Tr}}}

\newcommand{\expct}[1]{\left\langle #1 \right\rangle}

\def\be{\begin{equation}}
\def\ee{\end{equation}}
\def\bc{\begin{center}}
\def\ec{\end{center}}

\def\r2{{\sqrt{2}}}

\def\tr{{\rm tr}}

\def\bea{\begin{eqnarray}}
\def\eea{\end{eqnarray}}

\definecolor{burntorange}{rgb}{0.8, 0.33, 0.0}

\newcommand{\doublewidetilde}[1]{{%
  \mathpalette\double@widetilde{#1}%
}}
\newcommand{\double@widetilde}[2]{%
  \sbox\z@{$\m@th#1\widetilde{#2}$}%
  \ht\z@=.9\ht\z@
  \widetilde{\box\z@}%
}

\usepackage{amsfonts}\usepackage{tabularx}\usepackage{dcolumn}\usepackage{bm}\usepackage{graphicx}\usepackage{epstopdf}

\hypersetup{urlcolor=blue}
\usepackage{xr}
\usepackage{lipsum}
\usepackage{tensor}
\usepackage{braket}

\usepackage[capitalise,compress]{cleveref}
\crefname{section}{Sec.}{Secs.}
\Crefname{section}{Section}{Sections}
\crefrangelabelformat{equation}{\textup{(#3#1#4)}--\textup{(#5#2#6)}}

\makeatother

\usepackage{xcolor}

\definecolor{darkgreen}{rgb}{0.0, 0.6, 0.13}

\newcommand{\nz}{n^{(0)}}
\newcommand{\nza}{\nz_\alpha}

\newcommand{\Cee}{C^{\varepsilon\varepsilon}}

\usepackage{amsfonts}\usepackage{tabularx}\usepackage{dcolumn}\usepackage{bm}\usepackage{graphicx}\usepackage{epstopdf}

\hypersetup{urlcolor=blue}

\usepackage{xr}
\usepackage{xr-hyper}
\usepackage{hyperref}
\makeatletter
\newcommand*{\addFileDependency}[1]{
  \typeout{(#1)}
  \@addtofilelist{#1}
  \IfFileExists{#1}{}{\typeout{No file #1.}}
}
\makeatother

\makeatother
\begin{document}

\title{Energy diffusion in weakly interacting chains with fermionic dissipation-assisted operator evolution}
\author{En-Jui Kuo}
\affiliation{Department of Physics, University of Maryland, College Park, MD 20742, USA}
\affiliation{Joint Quantum Institute, NIST/University of Maryland, College Park, MD 20742, USA}
\affiliation{Joint Center for Quantum Information and Computer Science, University of Maryland, College Park,
Maryland 20742, USA}
\author{Brayden Ware}
\affiliation{Joint Center for Quantum Information and Computer Science, University of Maryland, College Park,
Maryland 20742, USA}
\author{Peter Lunts}
\affiliation{Department of Physics, Harvard University, Cambridge MA 02138, USA}
\affiliation{Joint Quantum Institute, NIST/University of Maryland, College Park, MD 20742, USA}
\author{Mohammad Hafezi}
\affiliation{Department of Physics, University of Maryland, College Park, MD 20742, USA}
\affiliation{Joint Quantum Institute, NIST/University of Maryland, College Park, MD 20742, USA}
\affiliation{Joint Center for Quantum Information and Computer Science, University of Maryland, College Park,
Maryland 20742, USA}
\author{Christopher David White}
\affiliation{Joint Center for Quantum Information and Computer Science, University of Maryland, College Park,
Maryland 20742, USA}

\begin{abstract}
Interacting lattice Hamiltonians at high temperature generically give rise
to energy transport governed by the classical diffusion equation; however, predicting the rate of diffusion requires numerical simulation of the microscopic quantum dynamics.
For the purpose of predicting such transport properties, computational time evolution methods must be paired with schemes to control the growth of entanglement to tractably simulate for sufficiently long times.
One such truncation scheme---dissipation-assisted operator evolution (DAOE)---controls entanglement by damping out components of operators with large Pauli weight.
In this paper, we generalize DAOE to treat fermionic systems.
Our method instead damps out components of operators with large fermionic weight. We investigate the performance of DAOE, the new fermionic DAOE (FDAOE), and another simulation method, density matrix truncation (DMT), in simulating energy transport in an interacting one-dimensional Majorana chain.
The chain is found to have a diffusion coefficient scaling like interaction strength to the fourth power, contrary to naive expectations based on Fermi's Golden rule---but consistent with recent predictions based on the theory of \emph{weak integrability breaking}. In the weak interaction regime where the fermionic nature of the system is most relevant, FDAOE is found to simulate the system more efficiently than DAOE.
\end{abstract}
\maketitle

\section{Introduction}

Simulating transport in strongly interacting systems is a core challenge in quantum many-body physics,
with implications from strange metal physics in cuprates and iron pnictides 
\cite{zaanen2019planckian,ayres2021incoherent,poniatowski2021counterexample,spivak2010colloquium,kasahara2010evolution,sachdev2011quantum,lucasMemoryMatrixTheory2015}
to heavy-ion collisions. \cite{stephanov_signatures_1998,kolb_anisotropic_2000,ollitrault_anisotropy_1992,star_collaboration_experimental_2010,heinz_exploring_2015}
Because complete numerical solution of a particular Hamiltonian is generally feasible only for small systems, 
transport simulations rely on approximate numerical methods.
But transport is understood in terms of two largely separate languages, depending on the degree of interaction:
nearly free fermion\cite{pitaevskiiPhysicalKineticsVolume1981} (and nearly Bethe ansatz integrable\cite{starkKineticDescriptionThermalization2013,bertiniPrethermalizationThermalizationModels2015,mallayyaPrethermalizationThermalizationIsolated2019,friedmanDiffusiveHydrodynamicsIntegrability2020,durninNonEquilibriumDynamicsWeakly2021,lopez-piqueresHydrodynamicsNonintegrableSystems2021,denardisStabilitySuperdiffusionNearly2021a}%
) systems can be understood in terms of Boltzmann theory
while strongly interacting systems are understood in terms of an increasingly detailed theoretical understanding of how thermalization and hydrodynamics emerge from unitary microscopic dynamics.
\cite{khemaniOperatorSpreadingEmergence2018,kvorningTimeevolutionLocalInformation2021,vonkeyserlingkOperatorBackflowClassical2022,whiteEffectiveDissipationRate2023,artiacoEfficientLargeScaleManyBody2023}
Cold atom experiments highlight this gap: they can tune from free-fermion to strongly interacting by tuning a Feschbach resonance\cite{brownBadMetallicTransport2019a,yanTwodimensionalProgrammableTweezer2022} or changing the geometry of a  quasi-1D ladder geometry.\cite{wienandEmergenceFluctuatingHydrodynamics2023}
At the same time, progress in analytical and numerical treatment of systems showing Bethe ansatz integrability,\cite{bertiniFinitetemperatureTransportOnedimensional2021a} weakly broken Bethe ansatz integrability,\cite{starkKineticDescriptionThermalization2013,bertiniPrethermalizationThermalizationModels2015,mallayyaPrethermalizationThermalizationIsolated2019,friedmanDiffusiveHydrodynamicsIntegrability2020,durninNonEquilibriumDynamicsWeakly2021,lopez-piqueresHydrodynamicsNonintegrableSystems2021,denardisStabilitySuperdiffusionNearly2021a}
and strong integrability-breaking interactions%
\cite{whiteQuantumDynamicsThermalizing2017,rakovszky2022dissipation,vonkeyserlingkOperatorBackflowClassical2022,whiteEffectiveDissipationRate2023,kvorningTimeevolutionLocalInformation2021,artiacoEfficientLargeScaleManyBody2023}
suggest that quantum simulation may not be necessary, at least for one-dimensional systems.
But classical methods have not been shown to work in the crossover regime between weak interaction, tractable with Boltzmann methods, and strong interaction, tractable with recent methods.
We present a matrix product operator method for simulating transport in one dimensional high-temperature quantum systems that is suitable for that regime; it can treat both nearly-free-fermion and strongly interacting Hamiltonians.

\begin{figure}[t]
    \centering
    \includegraphics[width=0.95\columnwidth]{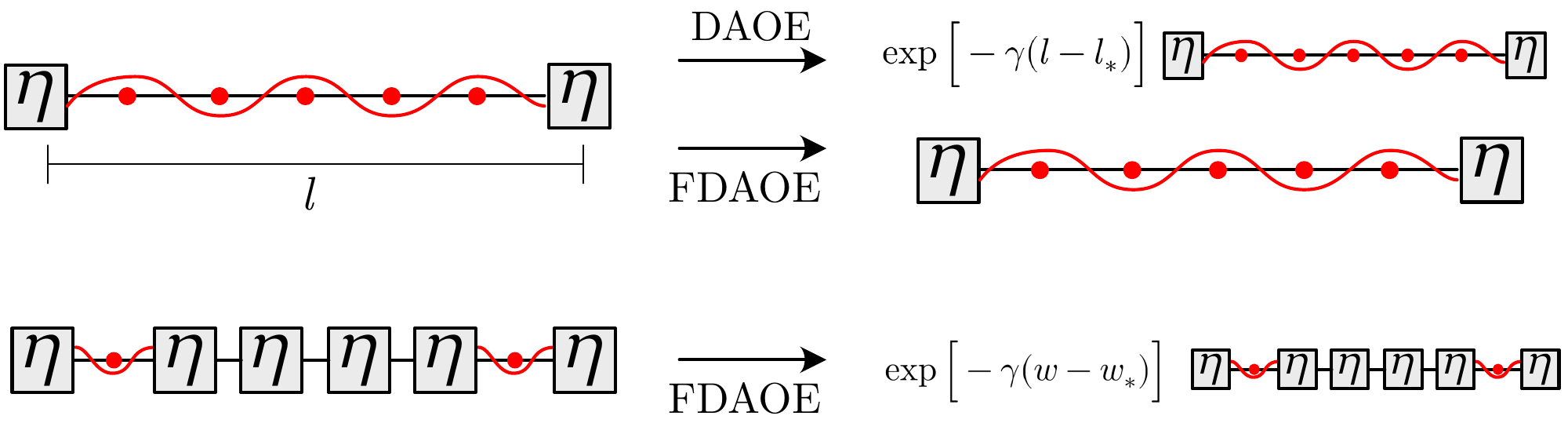}
    \vskip 0.25cm
    \includegraphics[width=0.95\columnwidth]{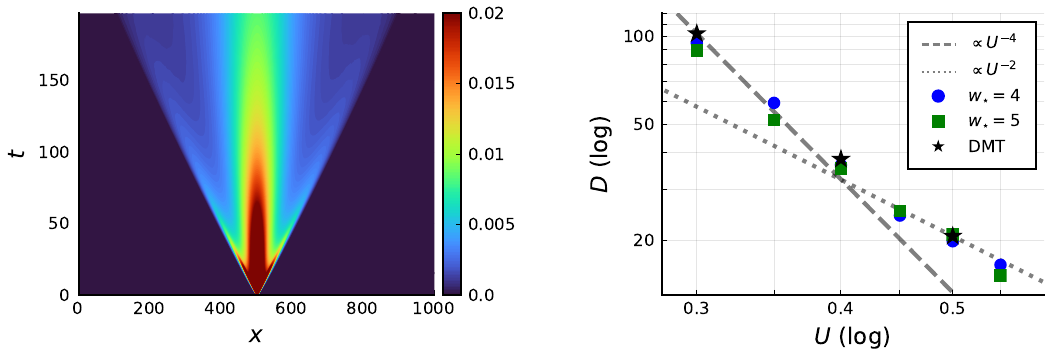}
    \caption{\textbf{Top}: dissipation-assisted operator evolution (DAOE) compared with fermionic dissipation-assisted operator evolution (FDAOE).
    Where DAOE reduces widely-separated quadratic fermion operators almost to zero, FDAOE leaves them untouched;
    FDAOE depolarizes operators with fermion weight greater than some cutoff $w_*$.
    \textbf{Bottom left:} energy density as a function of position $x$ and time $t$ in the nearly-free Majorana model \eqref{eq:ham-spin} at $U = 0.3$, simulated in FDAOE.
    \textbf{Bottom right:} finite-time estimates of the diffusion coefficient $D$. At small $U$, these estimates are consistent with $D \propto U^{-4}$, but not $D \propto U^{-2}$; the power is due to weak integrability breaking.
    }
    \label{fig:main}
\end{figure}

Existing methods for strongly interacting systems become impractical at weak interaction,
lack a perturbatively small simulation parameter controlling deviation from the exact dynamics, or both.
Density matrix truncation (DMT)\cite{whiteQuantumDynamicsThermalizing2017} works in all ranges of integrability,%
\cite{yeEmergentHydrodynamicsNonequilibrium2020,weiQuantumGasMicroscopy2022,yeUniversalKardarParisiZhangDynamics2022b,thomasComparingNumericalMethods2023}
but it is nontrivial to implement and difficult to analyze.
It is also uncontrolled:
like many matrix product operator methods, one checks the accuracy of DMT simulations by looking for convergence in bond dimension,
but for large systems, practical bond dimensions cannot approach the bond dimensions required to exactly simulate the state.
Indeed the premise of DMT, applied to systems that thermalize, is that most of the operator can be discarded, because it consists of physically-irrelevant correlations.

Dissipation-assisted operator evolution (DAOE) \cite{rakovszky2022dissipation} offers a controllable
approximation to a system's dynamics with a straightforward matrix product operator implementation---but it is not suitable for systems near free-fermion integrability.
DAOE modifies the Heisenberg dynamics to include an artificial dissipation-like superoperator with a parametrically small rate $\gamma$.
Just as depolarizing noise with rate $\gamma$
reduces the amplitude of operators with Pauli weight (number of nontrivial Pauli strings) $l$ at a rate  $\gamma l$,
the artificial dissipation
reduces the amplitude of operators with Pauli weight $l > l_*$ at a rate  $\gamma (l-l_*)$.
Fig.~\ref{fig:main} top shows a cartoon of this process.
It therefore reduces the state's complexity by decreasing the amplitude of long operators,
without changing local operators.
Because the long operators on which DAOE acts most strongly are%
---for chaotic, strongly interacting systems---%
unimportant to the finite-time dynamics\cite{vonkeyserlingkOperatorBackflowClassical2022},
one can think of the DAOE dissipation superoperator as perturbatively modifying the local dynamics.
DAOE results at small but finite $\gamma$ can therefore be extrapolated to the unitary $\gamma = 0$ dynamics of interest, in a manner similar to zero-noise extrapolation.\cite{liEfficientVariationalQuantum2017,temmeErrorMitigationShortDepth2017a} 

But when the system is not strongly interacting,
the high-weight operators affected by DAOE can be important to local dynamics,
and the dissipation is not a small perturbation.
In such a system momentum occupation numbers like $c^\dagger_k c_k$ (where $c_k$ is a fermion momentum mode annihilation operator)
are nearly conserved quantities,
so modifying them renders any description of the system's hydrodynamics unfaithful.
But simple long-range fermion operators like $c^\dagger_k c_k$ have large weight when written in Pauli matrices,
due to Jordan-Wigner strings,
so the artificial dissipation causes them to decay rapidly;
this artificial decay will dominate the system's apparent transport properties.

We modify DAOE to respect the fermionic structure underlying weakly interacting 1D systems;
we call the resulting method \textit{fermionic DAOE} (FDAOE).
FDAOE preserves quadratic fermion operators like $c^\dagger_k c_k$ while dissipating operator components consisting of products of large numbers of fermion operators.
Like DAOE it is efficient and easy to implement due to its compact matrix product operator representation.
This allows us to study both strongly and weakly non-integrable fermionic models using the same controlled and intuitive method.

We test FDAOE and a prior method, DMT, on a model displaying \textit{weak integrability breaking}. \cite{suraceWeakIntegrabilityBreaking2023}
In such models an integrability-breaking perturbation is added to an integrable (in our case free fermion) model.
At leading order, the perturbation dresses the integrable system's conserved quantities,
giving ballistic transport;
beyond leading order it scatters those quantities, giving diffusive transport.
Ref.~\onlinecite{suraceWeakIntegrabilityBreaking2023} predicts relaxation times $\sim U^{2+2\nu}$, where $\nu$ is a positive integer for perturbations that exhibit weak integrability breaking and $0$ for perturbations that do not.

We find that both FDAOE and DMT capture infinite-temperature dynamical correlation functions of energy density in such a system on short and intermediate timescales.
Both methods are limited by rapid growth of the patch of the system they must simulate:
on timescales short compared with the scattering time, which is itself not short,
energy density spreads nearly ballistically and one must simulate systems of diameter $\propto vt$.
Although FDAOE and DMT both allow simulation at bond dimensions $\sim 64-128$,
this spread still gives cost per timestep $\sim vt$ and total simulation cost $\sim t^2$.
FDAOE is additionally limited by SVD error.

Both methods give finite-time energy density diffusion coefficients consistent with $D \sim U^{-4}$ but not the simple Fermi golden rule $D \sim U^{-2}$ expected from ordinary integrability breaking.
This $U^4$ scaling confirms that weak integrability breaking governs the system's  dynamics
not only at times short compared to interaction and hopping,
where \onlinecite{suraceWeakIntegrabilityBreaking2023} worked,
but on long times as well. 

The paper is organized as follows.
In Sec.~\ref{s:model} we discuss the model used for benchmarking,
give a brief overview of weak integrability breaking,
and discuss the quantities of interest.
In Sec.~\ref{s:method} we review DAOE,
and present our new method FDAOE
and describe simulations and simulation costs.
In Sec.~\ref{s:results} we present the results and discuss the performance of FDAOE compared to DAOE,
and in Sec.~\ref{s:discussion} we conclude.

\section{Model and quantities of interest}\label{s:model}

\subsection{Model}

We study the infinite-temperature energy transport of an interacting Majorana chain
\begin{align}\label{eq:ham-maj}
    H = \sum_n i\eta_n \eta_{n+1} - U \sum_n \eta_{n-1} \eta_n \eta_{n+1} \eta_{n+2}\;.
\end{align}
We work in the units where the quadratic ``hopping" term has been set to one. We chose this model by starting with the simplest example of a 1D free-fermion model with energy conservation but without particle number conservation and adding the most natural fermion parity-conserving interaction. 
(The low-energy properties of this Hamiltonian were previously studied in Ref.~\onlinecite{rahmani2015phase}.)

This Hamiltonian is equivalent to the spin-1/2 Hamiltonian

\begin{equation}\label{eq:ham-spin}
    H = \sum_n \sigma^x_n\sigma^x_{n+1}+\sigma^z_n +U \left(\sigma^x_n\sigma^x_{n+2}+ \sigma^z_n\sigma^z_{n+1}\right).
\end{equation}
by Jordan-Wigner transformation.
We work with the spin-language Hamiltonian \eqref{eq:ham-spin}.

We choose a model \eqref{eq:ham-maj} with only a single conserved quantity,
the energy density,
so we can study transport in the simplest possible setting.
We analyze our simulations through the lens of a single-component diffusion equation.%
\cite{leviatanQuantumThermalizationDynamics2017,kvorningTimeevolutionLocalInformation2021,parkerUniversalOperatorGrowth2019,whiteEffectiveDissipationRate2023,rakovszky2022dissipation,thomasComparingNumericalMethods2023}

We do not expect our simulation methods to break down in systems with multiple conserved quantities.
But when a system has multiple conserved quantities,
non-linear interactions between the conserved quantities
can  contribute significantly to transport properties, so 
going beyond the linear diffusion equation is necessary to analyze such systems \cite{mukerjeeStatisticalTheoryTransport2006}.
Non-linear effects can also appear in systems with a single conserved quantity,\cite{mukerjeeStatisticalTheoryTransport2006,delacretazHeavyOperatorsHydrodynamic2020, michailidisCorrectionsDiffusionInteracting2023}
but previous numerical simulations in spin chains with only energy conservation
have not detected significant non-linear contributions to energy transport.\cite{thomasComparingNumericalMethods2023}
In this work we also do not detect significant non-linear contributions.

We have chosen an interaction that is not integrable using the Bethe ansatz.
Bethe ansatz integrable systems have an infinite hierarchy of additional conserved quantities beyond energy density.
The methods we use throughout this paper are designed to truncate irrelevant information while preserving the behavior of densities of conserved quantities, which are short-range or few-fermion operators like the energy density.
So we would not expect them to be appropriate for Bethe-ansatz integrable systems which have conserved quantities with arbitrarily large operator size: see Ref.~\onlinecite{rakovszky2022dissipation} App.~B and Ref.~\onlinecite{yooOpensystemSpinTransport2023} for more discussion of this point.

While the Hamiltonian we study in this paper has no additional conserved quantities for non-zero $U$, we wish to study transport in a regime of small $U$, in the neighborhood of the free-fermion point. At this free point, the model has many additional conserved quantities that are quadratic in the fermion operators. Such quantities are almost conserved when $U$ is small but non-zero, and thus may contribute significantly to long-time dynamics. Prior studies using tensor network algorithms have not attempted to predict transport properties in the nearly free-fermion regime of 1D chains, and it is not clear whether such methods would succeed for this purpose. On the other hand, FDAOE is designed specifically to preserve information about the low-fermion weight quantities that are almost conserved in the small $U$ regime.

\subsection{Weak integrability breaking}

In the regime of weak interactions, transport can sometimes be studied using simpler means, with the transport coefficients computed perturbatively in linear response using the Kubo formula, avoiding the need for simulations of operator  dynamics.

However, we expect that the model studied in this paper evades a 

simple perturbative analysis, as it exhibits \textit{weak integrability breaking}.\cite{surace2023weak} The core prediction of weak integrability breaking for this model is that scattering times do not scale as predicted by Fermi's golden rule, with the square of the perturbation strength $U$, but with a different power-law scaling. This is due to a hidden non-local map that approximately transforms the Hamiltonian into a free fermion Hamiltonian, which we describe below. This implies that the in the perturbative calculation current-current correlators need to be computed to a higher order than expected to obtain a finite result (in this case fourth order rather than second order in $U$). The simulation methods in this paper are able to recover the unusual scaling with $U$ without any explicit knowledge of the weak integrability breaking.

Weak integrability breaking starts from the elementary observation that
if $n^{(0)}_\alpha$ is a conserved quantity of some Hamiltonian $H_0$,

then 
\begin{align}
n'_\alpha = e^{i\lambda X} \nza e^{-i\lambda X}\;,
\end{align}
where $X$ is an Hermitian operator, is a conserved quantity of
\begin{align}
    H' = e^{i\lambda X} H_0 e^{-i\lambda X}\;.
\end{align}
But for small $\lambda$, this $H'$ is not so far from
\begin{equation}\label{eq:general-dressed-ham}
    H = H_0 + i\lambda [X, H_0] = H' + O(\lambda^2)\;.
\end{equation}
When the initial Hamiltonian $H_0$ is chosen to be an integrable Hamiltonian with conserved quantities $\{n^{(0)}_\alpha \}$, the perturbed quantities
\begin{equation}\label{eq:general-dressed-cons}
   n_\alpha = n^{(0)}_\alpha + i\lambda [X,n^{(0)}_\alpha] = n'_\alpha + O(\lambda^2)
\end{equation}
are nearly conserved.
That is, 
\begin{equation} \label{eq:general-matrix-element}
    [H, n_\alpha] \propto \lambda^2\;,
\end{equation}
in contrast to a generic perturbation which leads to matrix elements $[H_0 + V, \nza] \propto \lambda$\;.
Indeed by keeping higher commutators in \eqref{eq:general-dressed-ham}, \eqref{eq:general-dressed-cons},
one can engineer nearly conserved quantities with arbitrarily slow decays $[H, n_\alpha] \propto \lambda^{2k}$.

The challenge is to find generators $X$ which produce a local perturbation $V$ of interest via $V = i [X, H_0]$, which is only possible for select perturbations $V$.
Ref.~\onlinecite{surace2023weak} systematically constructs a family of non-local generators $X$
that produce local perturbations when the unperturbed Hamiltonian $H_0$ is free-fermion or Bethe ansatz integrable.
Specifically, the examples of generators $X$ that they construct are bi-local combinations of conserved densities $n^{(0)}_\alpha$ of $H_0$ and their corresponding current operators.

For our purposes, we only need to consider one such generator $X$ for a free Majorana chain that produces the interaction term of \cref{eq:ham-maj}.
The appropriate generator is constructed from a non-local combination of the energy density and energy current density operators of the unperturbed Hamiltonian:
\begin{align}
\begin{split}
    X &= \sum_{m<n} \{\varepsilon_m^{(0)}, j_n^{(0)}\} +  \frac 1 2 \sum_n \{\varepsilon_n^{(0)}, j_n^{(0)}\}\\
  \varepsilon_n^{(0)} &= \sigma^x_n \sigma^x_{n+1} + \frac 1 2 ( \sigma^z_n + \sigma^z_{n+1})\\
  j_n^{(0)} &= \sigma^x_n \sigma^y_{n+1} - \sigma^y_n\sigma^x_{n+1}\;.
\end{split}
\end{align}
One can check (with some purely mechanical effort) that the interacting Hamiltonian in \cref{eq:ham-maj} satisfies
\begin{equation}
    H = H_0 + i\lambda [X,H_0]
\end{equation}
with
\begin{subequations}
\begin{align}
    H_0 &= \sum_j i \eta_j \eta_{j+1} \\
\end{align}
and $\lambda = U/4$.
\end{subequations}

\subsection{Quantities of interest}\label{s:quantities}

If a system thermalizes, it approaches local thermal equilibrium:
after a short time, it can be described by a density matrix 
\begin{align} \label{eq:local-eq}
    \rho(t) \propto \exp\left[-\sum_x \beta_x(t) \varepsilon_x\right]\;,
\end{align}
where $\beta_x(t)$ is a smoothly varying space- and time-dependent inverse temperature and $\varepsilon_x$ is the energy density.
(We consider systems with only one conserved quantity, the energy density; a discussion of systems with more than one conserved quantity would proceed analogously.)
This state is specified entirely by the energy expectation values $\tr \rho(t) \varepsilon_x$.
For times longer than the initial thermalization time the energy density correlation function
\begin{align}
\begin{split}
    C(x,t) = \expct{\varepsilon_x(t)\varepsilon_0(t)}\;
\end{split}
\end{align}
therefore captures the whole dynamics in this long-time regime,
and the extent to which it deviates from a gradient-expansion prediction from \eqref{eq:local-eq}
diagnoses the local thermalization process.

On timescales long compared with the local thermalization time, the system's dynamics are given by the continuity equation and a gradient expansion of the state \eqref{eq:local-eq};
the result is
\begin{align} \label{eq:fick}
\begin{split}
    \partial_t \varepsilon &= \partial_x j\\
    \partial_t j &= D \partial_x \varepsilon + \dots.
\end{split}
\end{align}
To the extent that the system is described by the leading-order term in the gradient expansion \eqref{eq:fick},
the energy density correlation function is the Gaussian $C(x,t) \propto \exp\left[- x^2/4Dt\right]$.

But  real systems are only described by \eqref{eq:fick} on timescales long compared to the microscopic thermalization timescale.
To characterize the correlation function $C(x,t)$ at short or intermediate timescales,
we can measure the degree to which it spreads away from the initial point $x = 0$ to introduce a time-dependent diffusion coefficient.
The degree of spread is the mean squared displacement
\begin{align}
    V(t) = \frac 1 \nu \left[ \sum_x x^2 C(x,t) - \left(\sum_x x C(x,t)\right)^2 \right]
\end{align}
where $\nu$ is a (time-independent) normalization
\begin{align}\label{eq:app:C-norm}
    \nu = \sum_{x} C(x,t) = \sum_x C(x,0) = \expct{\varepsilon_{L/2}^2}\;.
\end{align}
The time-dependent diffusion coefficient is
\begin{align}\label{eq:Dt-def}
    D(t) = \frac 1 2 \frac{d}{dt} V(t).
\end{align}
If the system is diffusive, then in the long-time limit its dynamics approach the diffusion equation \eqref{eq:fick} with
\begin{align}
    D = \lim_{t \to \infty} D(t)\;.
\end{align}
We estimate $D(t)$ by a numerical derivative of the mean squared displacement $V(t)$.

\section{Method: Fermion dissipation-assisted operator evolution}\label{s:method}

\subsection{Intuition and superoperator} \label{s:superoperator}
Dissipation-assisted operator evolution \cite{rakovszky2022dissipation}
intersperses unitary time evolution
with a dissipation superoperator that reduces the amplitude on high-weight Pauli strings.
That dissipation superoperator is
\begin{align}\label{eq:DAOE-D}
    \mathcal{D}_{l_{*},\gamma}[\mathcal{S}]=\left\{
    \begin{array}{lr}
     \mathcal{S} & \text{if } l_{\mathcal{S}} \leq l_{*}\\
     e^{-\gamma(l_{\mathcal{S}}-l_{*})}\mathcal{S} & l_{\mathcal{S}} > l_{*}
    \end{array}
\right\} \;,
\end{align}
where $\mathcal S$ is a Pauli string and $l_{\mathcal S}$ is the \textit{Pauli weight} of $\mathcal S$,
or the number of nontrivial Pauli operators in $\mathcal S$.
In the $l_* = 0$ limit, this reduces to a depolarizing channel, hence the name ``dissipation superoperator'';
from the point of view of DMT \cite{whiteQuantumDynamicsThermalizing2017} or operator size truncated dynamics,\cite{whiteEffectiveDissipationRate2023,thomasComparingNumericalMethods2023} it is a soft truncation on long operators.
When the dynamics of long operators is chaotic, the details of the dynamics of long operators does not affect local dynamical correlation functions \cite{vonkeyserlingkOperatorBackflowClassical2022,whiteEffectiveDissipationRate2023},
and for $l_* \gg 1$ the superoperator \eqref{eq:DAOE-D} modifies the local dynamics only by modifying the rate at which amplitude escapes from short operators to long operators.
Heuristically, the DAOE superoperator projects out operators with weight $l_{\mathcal S} \gtrsim l_* + \gamma^{-1}$.

But for many models of interest
the dynamics of long operators is not chaotic,
and the DAOE dissipation superoperator \eqref{eq:DAOE-D} dramatically changes operators of interest.
In a system of weakly interacting fermions like \eqref{eq:ham-maj}, a momentum mode such as 
\begin{align}\label{eq:k-mode}
    n^{(0)}_k = \frac 1 L \sum_{m<n} \eta_m \eta_n \sin (n - m)k
\end{align}
is a conserved quantity of the non-interacting part, and the system's evolution is governed by the dynamics of these momentum modes, together with scattering between them. 
After Jordan-Wigner transformation the term  $\eta_m \eta_n$ picks up a Jordan-Wigner string between $m$ and $n$,
so it has Pauli weight $l_{\eta_m \eta_n} = m-n+1$.
The DAOE superoperator projects out operators with Pauli weight $l_{\mathcal S} \gtrsim l_* + \gamma^{-1}$,
so it truncates the momentum occupation number to range $l_* + \gamma^{-1}$.
Because it changes the conserved quantities of the free model, one expects it to drastically change the transport properties of the interacting model.

In \textit{fermonic dissipation assisted operator evolution} (FDAOE)
we replace the Pauli weight $l_{\mathcal S}$ in the DAOE dissipation superoperator \eqref{eq:DAOE-D}
by a fermion weight superoperator.
To write the fermion weight superoperator,
first represent each Pauli matrix by two Majorana fermion operators
\begin{align}\label{eq:sigma-as-majorana}
\begin{split}
    \sigma^x_{n} &= \eta_{2n+1} \left[\prod_{m \geq n+1} i \eta_{2m} \eta_{2m+1}\right] \\
    \sigma^y_{n} &= \eta_{2n} \left[\prod_{m \geq n+1} i \eta_{2m} \eta_{2m+1}\right] \\
    \sigma^z_{n} &= -i\eta_{2n}\eta_{2n+1}\;.
\end{split}
\end{align}

These Majorana operators form a Hermitian, orthogonal basis for the space of operators,
so we can define our Majorana dissipation superoperator by its action on them:
if $\eta_J$ is a product of Majorana operators on the sites $J$
\begin{align}
    \eta_J = \prod_{j \in J} \eta_j\;,
\end{align}
then
\begin{align}\label{eq:FDAOE-M}
    \mathcal M_{w_*,\gamma}[\eta_J] 
    &=
    \begin{cases}
     \eta_J & w_{\eta_J} \le w_*\\
     e^{-\gamma(w_{\eta_J} - w_*)} \eta_J & w_{\eta_J} > w_*
    \end{cases}
\end{align}
with
\begin{align}
    w_{\eta_J} = |J|\;.
\end{align}
All of the terms in the momentum mode $n^{(0)}_k$ of \eqref{eq:k-mode}  have Majorana weight 2,
so for $w_* \ge 2$ and any $\gamma$
\begin{align}
    \mathcal M_{w_*,\gamma}(n_k^{(0)}) = n_k^{(0)}\;.
\end{align}
Unlike DAOE, then, FDAOE preserves the conserved quantities of the non-interacting Hamiltonian.

The FDAOE superoperator \eqref{eq:FDAOE-M} has an MPO representation with bond dimension  $w_* + 2$ when represented by its action on the Pauli basis related to the Majorana operators by the Jordan-Wigner transformation, \cref{eq:sigma-as-majorana}.
We give the MPO explicitly in Appendix \ref{app:FDAOE-mpo},
but outline the construction here.
As in the case of the DAOE superoperator described in Ref.~\onlinecite{rakovszky2022dissipation}, the MPO representation utilizes a constant rank-4 tensor which we denote as $W^{n n'}_{ab}$, with the upper indices $n, n'$ taking values in the local Pauli operator basis $\{I, X, Y, Z\}$. As the FDAOE superoperator is diagonal in the basis of operators that consist of products of Majorana operators, it is also diagonal in the basis of Pauli strings; thus, the only non-zero matrix elements occur in the form $W^{II}_{ab}$, $W^{XX}_{ab}$, $W^{YY}_{ab}$, or $W^{ZZ}_{ab}$. The virtual indices $a, b$ take values in the set  $\{0, 1, \ldots w^*-1\} \cup \{ (w^*, +), (w^*, -)\}$, which track the total fermionic operator weight as measured from the left end of the chain to the bond in question until it reaches $w^*$, and the fermion parity afterwards. Within each Pauli string, the additional fermionic weight represented by the presence of an $X$ or $Y$ is always $1$ and always flips the fermion parity, which is tracked by the virtual index. Consequently, $W^{XX}_{ab}$ and $W^{YY}_{ab}$ are zero unless $a<w^*$ and $b=a+1$ or $a=(w^*, \pm)$ and $b=(w^*, \mp)$. The weight associated with $I$ and $Z$ however are context dependent; alone, they correspond to weights $0$ and $2$ respectively but within a Jordan-Wigner string they correspond to weights $2$ and $0$, swapping roles. The presence of a Jordan-Wigner string is locally accessible as the parity of the MPO virtual index $a$. In addition to tracking the fermion weight, the MPO tensor applies a decay factor of $e^{-\gamma}$ for each unit of additional fermion weight beyond $w^*$. Finally, the tensors are contracted on the left and right ends with the vectors $v_L = (1, 0, \ldots, 0)$ and $v_R = (1, 1, \ldots, 1)$; this ensures the virtual index starts tracking the fermion weight from weight $0$ on the left end of the chain and allows all values of fermion weight on the right end.
Explicit expressions for the MPO tensors that meet these conditions are given in Appendix ~\ref{app:FDAOE-mpo}.

\subsection{Computing dynamical correlation functions}

We seek to measure dynamical correlation functions
\begin{equation}
   \expct{ \varepsilon_x(t) \varepsilon_0(0) } = \tr \left( \varepsilon_x e^{-i\mathcal L t} [\varepsilon_0] \right)\;,
\end{equation}
where $\mathcal L$ is the Liouvillian generated by $\mathcal L[\cdot] = -i[H,\cdot]$ and where
$\varepsilon_x$ is the energy density, chosen as a parity symmetric operator that produces $H$ when summed over sites. Explicitly, we represent $\varepsilon_x$ as
\begin{align}
\begin{split}
\varepsilon_x =
\frac{1}{3}&\left(\sigma^{z}_x+\sigma^{z}_{x+1}+\sigma^{z}_{x+2}\right) \\
+\frac{1}{2}&\left(\sigma^{x}_x \sigma^{x}_{x+1}+\sigma^{x}_{x+1} \sigma^{x}_{x+2} \right)\\
+ \frac{U}{2}&\left(\sigma^{z}_x \sigma^{z}_{x+1}+ \sigma^{z}_{x+1} \sigma^{z}_{x+2}\right) \\
+U &\left(\sigma^{x}_x \sigma^{x}_{x+2}\right)\;.
\end{split}
\end{align}
To measure $\expct{ \varepsilon_x(t) \varepsilon_0(0) }$,
we time-evolve the initial operator $\varepsilon_0$ by the Trotterization of that Liouvillian,
interspersed with applications of the FDAOE MPO.

In the limit of large bond dimension, the dominant cost is truncation after MPO application.
Since the FDAOE MPO has bond dimension $w_*+2$, an exact truncation has cost
\begin{align}
    \sim \left[d^2 \chi \cdot (w_*+2)\right]^3
\end{align}
per bond, where $d = 2$ is the physical on-site Hilbert space dimension.
This cost comes about because the exact truncation requires two sweeps,
the first to put $\mathcal M_{w_*,\gamma}\circ e^{-i\mathcal L t} [\varepsilon_0] $ in canonical form
and the second to do the truncation.
Switching to a so-called ``zip-up'' truncation, in which one truncates at each site immediately after applying the MPO tensor,
reduces the cost to
\begin{align}
    \sim (w_*+2) \left[d^2 \chi\right]^3
\end{align}
per bond
at the cost of some imprecision.\cite{Stoudenmire_2010, PAECKEL2019167998}

The cost of the whole calculation can further be reduced by noting that $e^{-i\mathcal L t} [\varepsilon_0] $ acts as the identity
outside a lightcone of diameter $\sim 2vt$ for some speed $v$.
Both the unitary dynamics and the FDAOE MPO $\mathcal M_{w_*}$ act trivially outside that lightcone.
So the cost of an MPO application at simulation time $t$ generically grows with $t$:
it is
\begin{align}
    \sim (w_*+2) d^6\sum_x \chi(x,t)^3 \sim d^6(w_*+2) \chi(t)^3 \cdot 2vt\;,
\end{align}
where we write $\chi(x,t)$ for the bond dimension at site $x$ and some site $t$, and $\chi(t)$ for a typical magnitude at time $t$.
The memory requirements are
\begin{align}
    \sim (2vt)\cdot d^6\chi(t)^2\;.
\end{align}

\begin{figure}[t]
    \centering
    \includegraphics[width=0.45\textwidth]{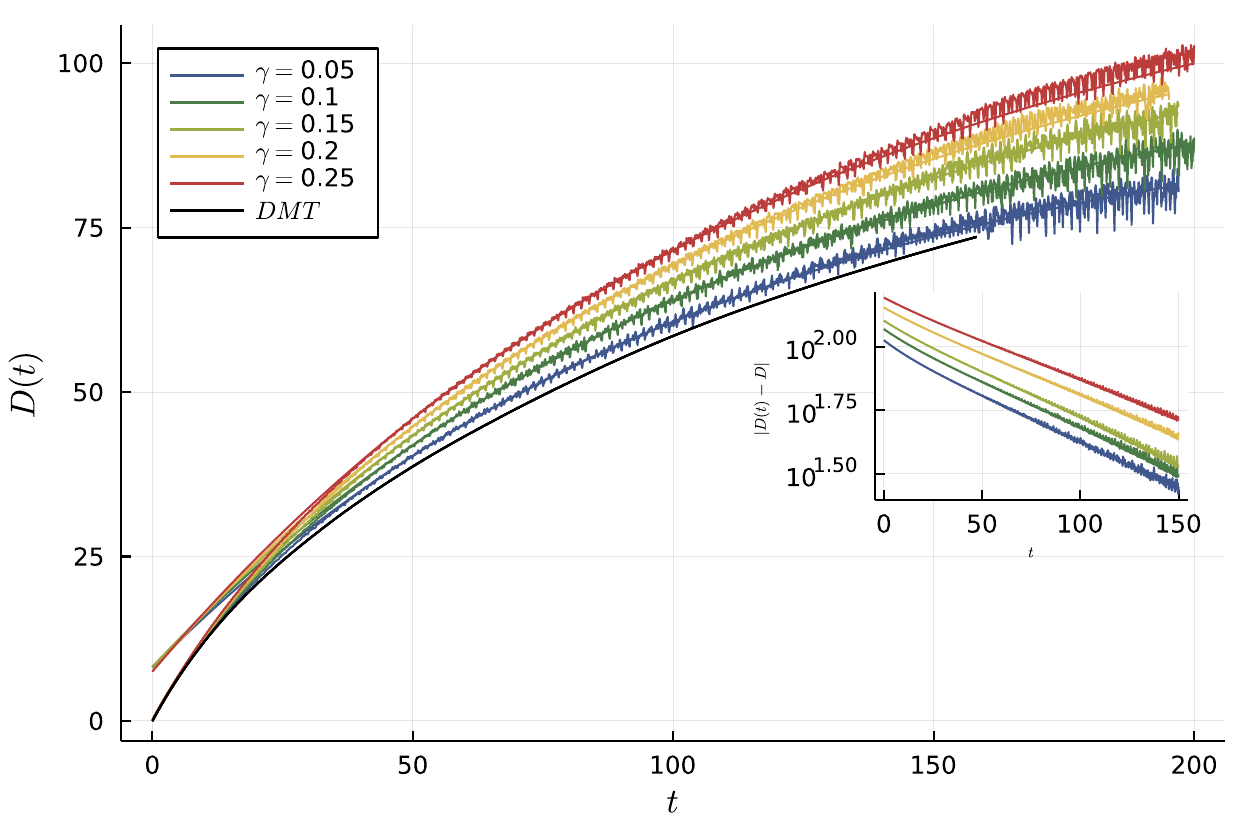}
    \caption{
    Diffusion coefficient $D(t)$ as a function of time for the nearly free Majorana model of Sec.~\ref{s:model} at interaction $U = 0.3$. $D(t)$ is estimated via Eq.~\ref{eq:Dt-def} using FDAOE at $w_* = 5$ and a range of artificial dissipations $\gamma$ (colored lines), and DMT at bond dimension 256 and Trotter step 0.125 (black line). (See App.~\ref{app:DMT} for DMT convergence data.) The noise in the FDAOE $D(t)$ is due to an SVD cutoff $\epsilon = 10^{-8}$ (see App.~\ref{app:epsilon-convergence}). For plots of other $U$, see Fig.~\ref{fig:D-gamma-other-U}.)
    } 
    \label{fig:Dt-03}
\end{figure}

\subsection{Simulation parameters}

We use a fourth-order Trotter decomposition; specifically, we use the three term formula recommended by Ref.~\onlinecite{barthel2020optimized}, that consists of $21$ layers of three site gates for each time step. The size of the time steps is of size $dt=0.1$ throughout the paper. The FDAOE MPO is applied after each time step.

After each Trotter gate and during zip-up MPO application we discard the smallest singular values $s_{\alpha}$ such that%
\footnote{ITensor svd keywords use\_relative\_cutoff=true, cutoff=1e-8}
\begin{align}
\sum_{\alpha\text{ discarded}} s_\alpha^2 < \epsilon \left[ \sum_\beta s_\beta^2\right]\;,
\end{align}
where $\epsilon$ is the SVD truncation error.

In App.~\ref{app:epsilon-convergence} we discuss convergence in the cutoff $\epsilon$ and the magnitude of the noise in the numerical derivative $D(t)$
(cf Fig.~\ref{fig:Dt-03}).
We empirically find that truncating singular values causes noise in $D(t)$ of magnitude
\begin{align}\label{eq:Dt-noise-pred-maintext}
    \Delta t_{\mathrm{pred}} \sim \sqrt{\epsilon}\, V(t)\;,
\end{align}
and give a heuristic argument for why the noise should have that form.
We also empirically find that the magnitude of the noise gives a reasonable estimate of the convergence error in $\epsilon$.

The DMT simulations are run at Trotter step $dt = 0.125$ and a fixed bond dimension cap $\chi = 256$; we discuss convergence in bond dimension (and other details of the DMT simulations) in App.~\ref{app:DMT}.

\section{Results}\label{s:results}

\begin{figure}[t]
    \centering
    \includegraphics[width=0.45\textwidth]{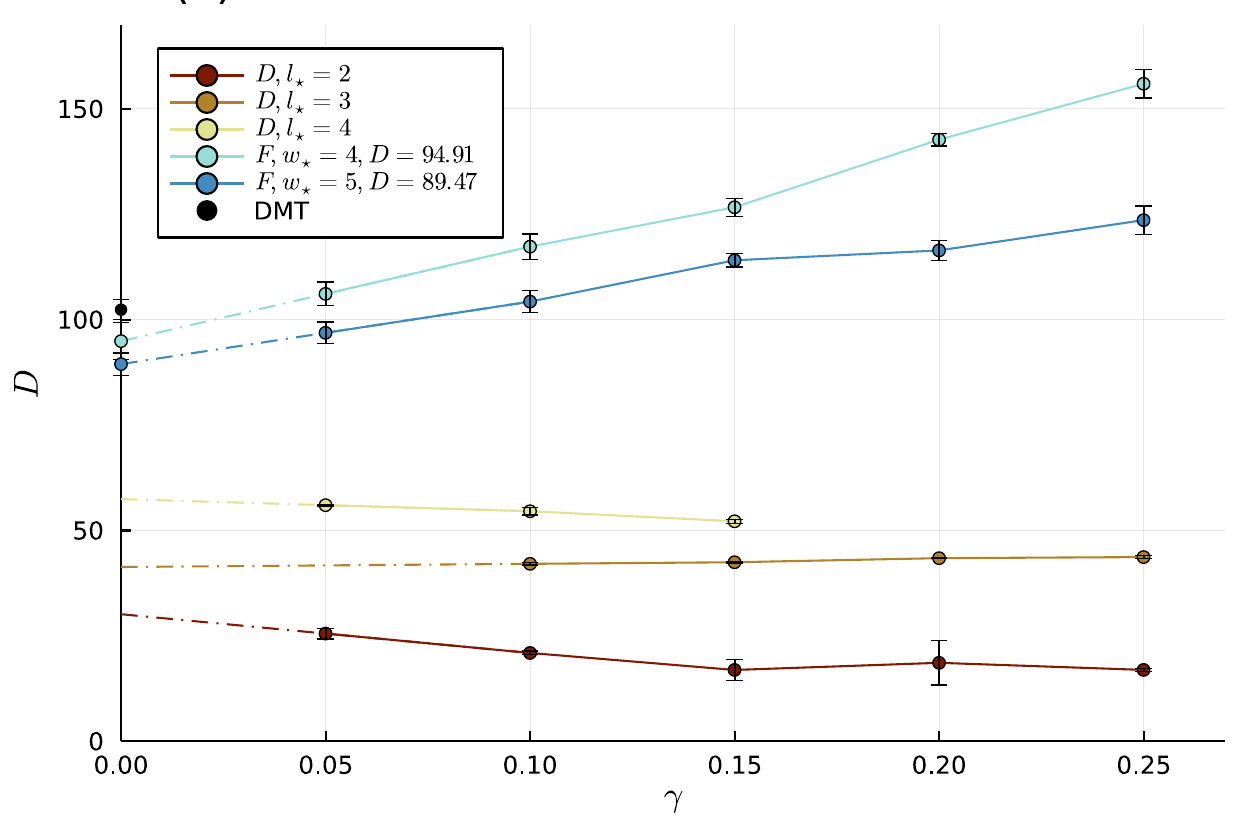}
    \caption{ Diffusion coefficients extracted from $D(t)$ via fit to exponential form \eqref{eq:Dt-fit-form} in FDAOE (blue lines, labeled ``F'' in legend) and DAOE (yellow-red lines, labeled ``D'' in legend). DAOE does not give consistent results across $l_*$, even when extrapolated to $\gamma = 0$,
    indicating that it cannot capture the crossover from free-fermion to diffusive dynamics. FDAOE gives consistent results across $w_*$; the diffusion coefficients extrapolated to $\gamma = 0$ are 
    $D_{w_* = 4} = 95$, $D_{w_* = 5} = 89$, against a DMT value $D_{\text{DMT} } = 102$. Error bars indicate uncertainty resulting from choice of fit window;
    they do not include uncertainty due to Trotter or truncation error.
    } 
    \label{fig:Dinf-gamma}
\end{figure}

Fig.~\ref{fig:Dt-03} shows $D(t)$ in the nearly free Majorana model of Sec.~\ref{s:model} at interaction $U = 0.3$,
estimated by taking the numerical derivative of FDAOE simulations of $\expct{\varepsilon_x(t)\varepsilon_0(0)}$.
The simulations are limited to times $t \lesssim 200$ by the computational cost, which grows with time as the lightcone of $\varepsilon_x(t)$ grows.
Each $D(t)$ shows high-frequency noise;
this is noise is controlled by the SVD truncation cutoff, here $\epsilon = 10^{-8}$ (cf App.~\ref{app:epsilon-convergence}.)

\begin{figure}
    \includegraphics[width=0.45\textwidth]{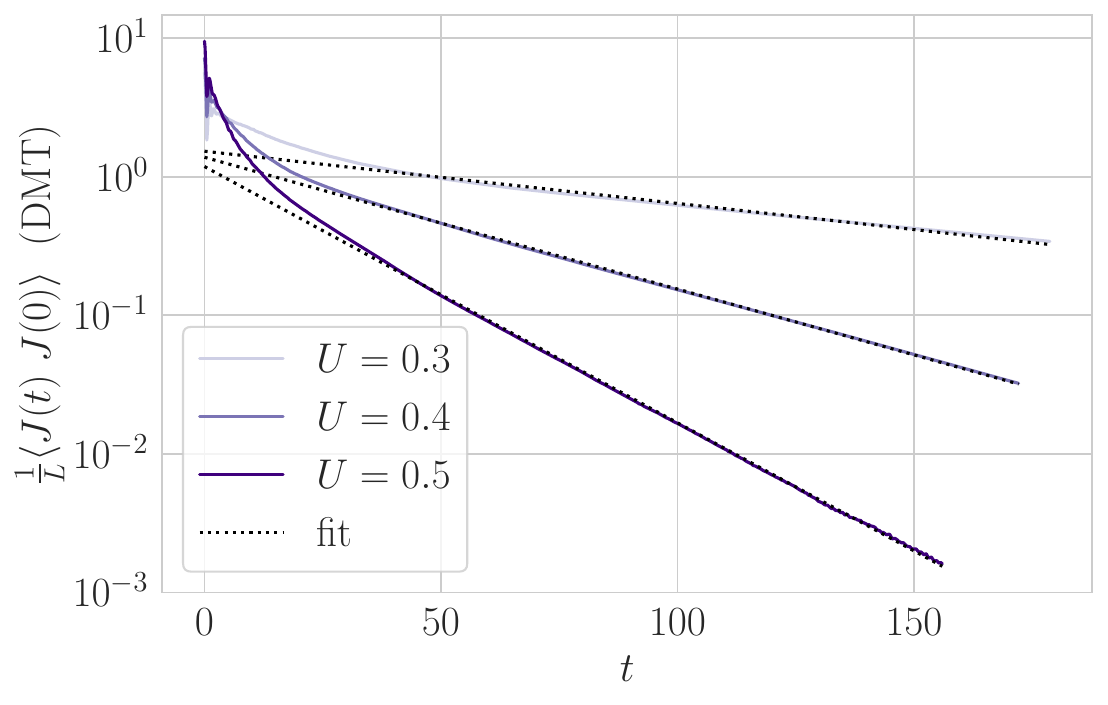}
    \caption{\textbf{Current-current correlator} in DMT simulations across interaction strengths $U$ (solid lines), together with a single-exponential fit to the $t > 25$ data.
    In each case the correlator displays a long-time exponential decay, supporting our choice of fit form \eqref{eq:Dt-fit-form} for the diffusion coefficient. (We cannot rule out power-law behavior at longer times, especially for $U = 0.3$, which displays less than a decade of decay over the time simulated.}
    \label{fig:dmt-JJ}
\end{figure}

After an initial transient behavior, $D(t)$ is described by exponential decay to a long-time limit
\begin{align} \label{eq:Dt-fit-form}
    D(t) = D - A e^{-t/\tau}\;.
\end{align}
Fig.~\ref{fig:Dt-03} inset shows $|D(t) - D|$ in FDAOE simulations, where $D$ is extracted by fitting FDAOE $D(t)$ to the form \eqref{eq:Dt-fit-form}.
The result appears exponential over the range of our simulations,
although that range is small (covering only a factor of about $3$ in decay of $|D(t) - D|$ for $U = 0.3$).
Fig.~\ref{fig:dmt-JJ} shows the current-current correlator
\begin{align}
 \frac 1 L \expct{J(t)J(0)} \propto \frac d {dt} D(t)
\end{align}
in DMT simulations for $U = 0.3, 0.4, 0.5$.
(See App.~\ref{app:DMT} for details of the DMT simulations, including convergence testing, Trotterization, and the definition of the current operator.
Note that that current operator corresponds to a different definition of energy density, 
for reasons of convenience in analytical calculations; this explains the difference in transient early-time behavior.)
For each $U$ the current-current correlator displays an early-time transient decay followed by a long-time exponential decay.
For $U = 0.5$ this decay covers approximately two orders of magnitude, but for $U = 0.3$ it covers less than a decade.
As in Ref.~\onlinecite{thomasComparingNumericalMethods2023} we cannot rule out that the system displays long-time tails, but we expect that if they exist their coefficients are small.

To characterize how the system's long but finite-time behavior depends on $U$ we fit to the form \eqref{eq:Dt-fit-form} for $U$-dependent time windows and consider the ``asymptotic'' diffusion coefficient $D$;
if the long-time tails are small or non-existent,
this $D$ will match the system's true diffusion coefficient.
We fit on time windows $15/U = t_{\min} \le t \le t_{\max} = 30/U $;
this form is chosen by eye to avoid both the early-time non-exponential behavior and late-time noise.
The window choice is fairly arbitrary.
To characterize how the window choice affects the fit,
we take end times $t_{\max} = 30/U - 10, 30/U, 30/U + 10$, fit for each window, compute the standard deviation of the three resulting diffusion coefficients,
and plot the result as error bars.

Fig.~\ref{fig:Dinf-gamma} shows the resulting diffusion coefficients for $U = 0.3$ as a function of the artificial decay $\gamma$.
We show both FDAOE and DAOE at a variety of $l_*, w_*$.
In each case we linearly extrapolate the smallest two points ($\gamma = 0.05, \gamma = 0.1$) to $\gamma = 0$.
We extrapolate each fit window separately; the point and error bar in the FDAOE extrapolation, like the point and error bar in each of the finite-$\gamma$ points, show the mean and standard deviation across the three fit windows.
The DAOE diffusion coefficients are not converged in $l_*$, in the sense that different $l_*$ give different extrapolations to $\gamma = 0$;
this indicates that DAOE is not in the perturbative small-$\gamma$ regime.

The FDAOE diffusion coefficients are converged in $w_*$ in the sense that the error bars in the extrapolation to $\gamma = 0$ overlap: the difference between $w_* = 4$ and $w_* = 5$ is less than the fit uncertainty.
In judging convergence it is important to note that 
in simulations of $\varepsilon_x(t)$, FDAOE with these two $w_*$ in fact leave the same operator Hilbert space untouched.
Because the Hamiltonian and the energy density are both fermion parity even,
$\varepsilon_x(t)$ is also fermion parity even, and the lowest-weight operators that suffer dissipation have weight $w = 6$, regardless of whether $w_* = 4$ or $w_* = 5$.
($w_* = 6$ simulations were prohibitively time consuming.)

The FDAOE diffusion coefficients also broadly agree with the DMT simulations (black dots in Fig.~\ref{fig:Dinf-gamma}) and in Fig.~\ref{fig:main}.)
The error bars do not all overlap, meaning the difference between DMT, FDAOE at $w_* = 4$, and FDAOE at $w_* = 5$ is not within fit uncertainty.
But the fit is not the only source of uncertainty;
DMT also has some error due to bond dimension convergence,
FDAOE has some error due to SVD cutoff,
and both methods have uncertainty due to (different) Trotter decompositions.
(See App.~\ref{app:other-interaction} for other interaction strengths and some convergence data, and App.~\ref{app:epsilon-convergence} and \ref{app:DMT} for discussions of convergence.)
The fit to $D(t)$ seems to be more more sensitive to convergence error than $D(t)$ itself.

Fig.~\ref{fig:Dinf-gamma} shows that the diffusion coefficient is approximately linear in $\gamma$ for $\gamma \lesssim 0.15$.
This, together with (broad) agreement between the different $w_*$ and agreement between FDAOE and DMT,
suggest that we can treat the FDAOE superoperator perturbatively,
although a rigorous justification for such a perturbative treatment is lacking.

\begin{figure}[t]
    \centering
    \includegraphics[scale=0.7]{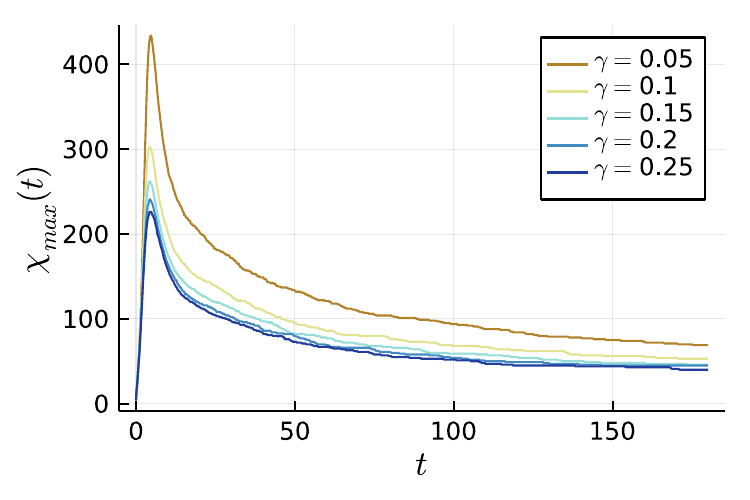}
    \caption{Bond dimension saturation plot with $w_{*}=5$ and $U=0.4$ for various $\gamma.$ We can see the bond dimension hits the peak and falls down. Here we pick the truncation error to be $10^{-8}$.}
    \label{fig:bd}
\end{figure}

The premise of FDAOE is that a weak projection on many-fermion operators will reduce simulation complexity,
while changing transport properties in a controlled way, which can safely be extrapolated to the limit of zero projection.
Fig.~\ref{fig:Dinf-gamma} establishes that the change in transport properties is controlled,
but not that simulation complexity is decreased.

Simulation complexity is controlled by bond dimension; Fig.~\ref{fig:bd} shows the maximum bond dimension as a function of time for $U = 0.4$, $w_* = 5$.
The bond dimension displays a fast initial rise followed by a slow decay,
as occurs in other examples of dissipative operator evolution.\cite{rakovszky2022dissipation, Noh2020efficientclassical}
The initial rise occurs as the operator $\varepsilon_x(t)$ spreads ballistically.
The long-time decay results from a split into conserved and non-conserved operators;
the non-conserved operators are destroyed by the FDAOE projection.
\cite{khemaniOperatorSpreadingEmergence2018}
Concretely, FDAOE turns the Heisenberg dynamics into an effective Liouvillian $\mathcal L_{\text{FDAOE}}$.
This effective Liouvillian has slow eigenoperators given by the Fourier transform of the energy density and the energy current:
\begin{align}
    \mathcal L_{\text{FDAOE}}[\tilde \varepsilon_k] = -iDk^2 \tilde \varepsilon_k\;,
\end{align}
with
\begin{align}
\begin{split}
    \tilde \varepsilon_k &= \varepsilon_k + ik Da j_k + \dots\; \\
    \varepsilon_k &= \sum_x e^{ikx} \varepsilon_x\\
    j_k &= \sum_x e^{ikx} j_x\;.
\end{split}
\end{align}
Here $a$ is a constant related to the normalization of $\varepsilon_k$ and $j_k$.
At long times, then, $\varepsilon_x(t)$ becomes a sum of these Fourier modes; performing an inverse Fourier transform it becomes
\begin{align}
    \varepsilon_0(t) = \sum_{x} \Big[\beta(x,t) \varepsilon_x + \eta(x,t) j_x + \dots \Big]\;,
\end{align}
with in fact $\eta(x,t) \propto D\partial_x \beta(x,t)$.
(This $\beta(x,t)$ is not identical to the inverse temperature of \eqref{eq:local-eq}, but it plays a similar role.)
At finite times this has small, decaying bond dimension;
in the long time limit $\lim_{t \to \infty} \partial_x \beta(x,t) = 0$, so the bond dimension will decay to that of the Hamiltonian.

We can rephrase the argument in the picture of \onlinecite{khemaniOperatorSpreadingEmergence2018}.
In that picture the initial energy density $\varepsilon_0$ develops weight on other energy densities $\varepsilon_x$ as it evolves,
but it also emits non-conserved, ballistically spreading operators.
In the exact unitary dynamics, these non-conserved operators would give large bond dimension.
But these operators not only spread spatially but also develop large fermion weight,
so FDAOE destroys them, leaving the energy density and its current.

\begin{figure}[t]
    \centering
    \includegraphics[scale=0.6]{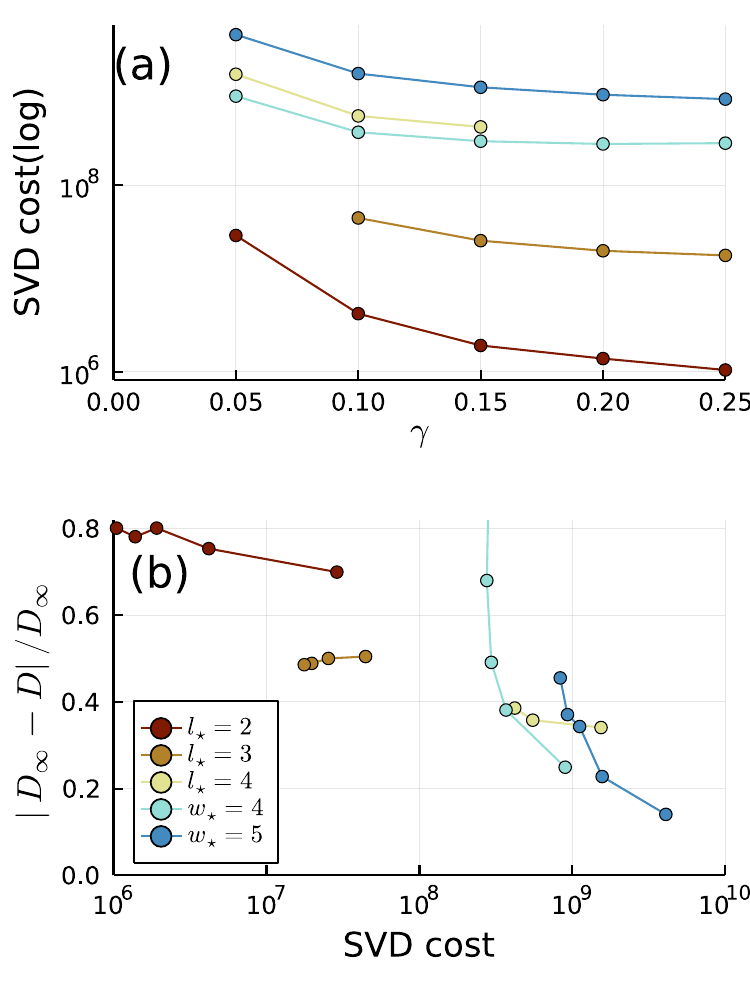}
    \caption{(a) SVD cost with various $\gamma$ for fixed $U=0.3$ and truncation error $10^{-8}$ (b) Relative error in the diffusion coefficient as a function of SVD cost. We can see that given a fixed cost, the FDAOE method has a faster convergence rate compared to DAOE.}
    \label{fig:svd}
\end{figure}

To measure performance in a hardware- and algorithm-agnostic way, we use the time complexity of SVD truncation after application of the superoperator MPO; we call this time complexity the \textit{SVD cost}.
Truncation dominates the time complexity of both DAOE and FDAOE,
so the SVD cost crudely estimates the number of floating point operations needed for the simulation.
Asymptotically, the time complexity of the SVD\cite{li2019tutorial} at bond $i$ is $\sim D\chi_i^3$, where $D$ is the bond dimension of the superoperator MPO: $D = l_* + 1$ for DAOE, and $D = w_* + 2$ for FDAOE.
For each method we sum over bonds at each time, and maximize over times:
\begin{align}
   \text{SVD cost} = \max_t \sum_i D \chi_i(t)^3 \;.
\end{align}
Fig.~\ref{fig:svd} shows the SVD cost for DAOE and FDAOE for $U = 0.3$ and cutoff $\epsilon = 0.3$.
Fig.~\ref{fig:svd} top plots SVD cost against the artificial dissipation rate $\gamma$;
it shows that---at any fixed $\gamma$---DAOE is one to two orders of magnitude cheaper.

But DAOE is not accurate:
recall that DAOE gave inconsistent diffusion coefficients between lengths $l_*$, and that none of those diffusion coefficients agreed with DMT.
FDAOE, by contrast, can be improved (at some cost in SVD time) by decreasing the artificial dissipation rate $\gamma$.
Fig.~\ref{fig:svd} bottom shows the error in the diffusion coefficient, compared to extrapolated FDAOE simulations, as a function of SVD cost.
Where DAOE plateaus at some large, $l_*$-dependent error,
FDAOE error decreases as the SVD cost increases.

\section{Discussion}\label{s:discussion}

We have presented a new method, fermionic dissipation assisted operator evolution (FDAOE),
for dynamics of interacting fermions in 1+1 dimensions at high temperature.
FDAOE modifies a previous method, dissipation assisted operator evolution (DAOE) \cite{rakovszky2022dissipation}, to perform a soft truncation of operators with large fermionic weight.
We tested FDAOE, DAOE, and another prior method, density matrix truncation (DMT),
on an interacting Majorana model displaying \textit{weak integrability breaking}.
In weak integrability breaking, interactions dress the conserved quantities of the free model,
leading to slow scattering and large diffusion coefficients.
We find that FDAOE and DMT agree and give diffusion coefficients consistent with the $D \sim U^4$ scaling ($U$ is the interaction strength), as expected from scattering of dressed fermions.
We further found that FDAOE decreases the bond dimension of the MPO representation of the Heisenberg picture energy density, when combined with SVD truncation with a small error cutoff,
but we also found that that small error cutoff induces noise in the time-dependent diffusion coefficient $D(t)$.

That SVD cutoff indirectly controlles the uncertainties in our diffusion coefficient estimates.
These uncertainties were controlled by uncertainty in the fit to the exponential form $D(t) = D - A e^{-t/\tau}$.
This fit uncertainty was driven in turn by truncation error,
because we stopped the fit where truncation error (estimated by noise in $D(t)$) became appreciable.
With less truncation error, we could measure $D(t)$ deeper into the exponential-decay regime,
improving our estimates of the asymptotic $D = \lim_{t \to \infty} D(t)$.
But most strategies for decreasing truncation error would impose substantial run-time and memory costs.
Reducing the SVD cutoff, e.g. from $10^{-8}$ to $10^{-9}$,
would directly increase the bond dimension throughout the simulation,
and replacing the SVD cutoff with a bond dimension cap
would prevent the simulation from taking advantage of the decay of the bond dimension (Fig.~\ref{fig:bd}).

But two strategies, measuring a current-current correlator and using a variable SVD cutoff---may reduce truncation error without imposing an appreciable runtime or memory cost.
Estimating $D(t)$ as we do, via the mean square displacement $V(t)$,
is costly because it requires measuring $\varepsilon(x,t)$ precisely at large $x$.
Indeed, the truncation error in $D(t)$ is $\propto \sqrt{\epsilon}\; V(t)$.
Measuring a current-current correlator would avoid this problem,
because the requisite spatial integral has no $x^2$
(unlike the integral giving $V(t)$).
Alternatively, one could use a time-varying SVD cutoff $\epsilon(t) = \epsilon_0 / V(t)^2$.
This would allow coarse simulations at early times, e.g. around the bond dimension peak in Fig.~\ref{fig:bd}, and fine simulations at later times.
Because the runtime cost contains large contributions from the bond dimension peak,
this might in fact decrease runtime cost on net.

While DMT is difficult to analyze,
FDAOE can be understood as a perturbative modification to the system's dynamics.
This is consistent with our observation that the FDAOE diffusion coefficient is linear in the artificial dissipation $\gamma$.
And the intuition is clear:
at small $\gamma$ FDAOE acts weakly on few-fermion operators, 
and  in systems that eventually thermalize,
many-fermion operators are not important to transport.
But establishing this formally will be nontrivial;
to do so will require modifying arguments like those of \onlinecite{vonkeyserlingkOperatorBackflowClassical2022}
to work away from the strongly-interacting, chaotic limit.

In one-dimensional fermion chains, the Jordan-Wigner transformation produces an equivalent local Hamiltonian that takes the form of a spin chain; a fermionic description of the transport is unnecessary in the generic case. For this reason, we have focused on weakly-interacting systems, where transport receives important contributions from the nearly conserved quadratic fermion operators of all sizes. While the Hamiltonian can be written as a local operator in the spin language, these nearly conserved operators cannot. By contrast, in higher dimensional fermionic systems, the energy density has no local spin representation. The dynamics of such systems can be computed using MPS by picking a one-dimensional ordering of the sites and using the Jordan-Wigner representation to convert to a spin Hamiltonian, with some terms involving non-local Jordan-Wigner strings. In this scenario, the DAOE superoperator will cause the energy density to decay rapidly, while the FDAOE superoperator will not. Thus, we expect that FDAOE is the appropriate choice for fermionic systems of all interaction strengths in higher dimensions. 

While we focused in this paper on infinite-temperature transport, extending the methods to allow for finite-temperature calculations is needed for many physical scenarios of interest. In scenarios where the equilibrium state is dominated by quadratic fermionic operators --- particularly at high temperatures with weak interactions --- the FDAOE operator will only weakly perturb the equilibrium. Thus, it may still be possible in these cases to recover the correct dynamics using FDAOE with the extrapolation to zero dissipation rate. We leave this question to future studies.

\textit{Note added}. -- We would like to bring the reader’s attention to a related independent work  \onlinecite{lloydBallisticDiffusiveCrossover2023}.

\section{Acknowledgements}
We would like to thank Federica Surace for helpful discussions about weak integrability breaking and sharing additional material related to Ref.~\onlinecite{surace2023weak}.
We wish to thank Stuart Thomas, Yong-Chan Yoo, Jay Deep Sau, and Brian Swingle for helpful conversations in the context of related collaborations.
E.J., M.H. were supported by W911NF2010232, AFOSR MURI FA9550-19-1-0399, Department of Energy QSA program (DE-AC02-05CH11231), and the Minta Martin and Simons Foundation. 
B.W.~was supported in part by the DoE ASCR Accelerated Research in Quantum Computing program (award No.~DE-SC0020312).
P.L. acknowledges support from the Simons Foundation, the Harvard Quantum Initiative Postdoctoral Fellowship in Science and Engineering, and the National Science Foundation under Grant No. DMR-2245246.
C.D.W. was supported by the U.S. Department of Energy (DOE), Office of Science, Office of Advanced Scientific Computing Research (ASCR) Quantum Computing Application Teams program, under fieldwork proposal number ERKJ347,
DOE Quantum Systems Accelerator program, DE-AC02-05CH11231,
AFOSR MURI FA9550-22-1-0339,
ARO grant W911NF-23-1-0242, ARO grant W911NF-23-1-0258, and NSF QLCI grant OMA-2120757.

\bibliography{sample}

\begin{thebibliography}{56}%
\makeatletter
\providecommand \@ifxundefined [1]{%
 \@ifx{#1\undefined}
}%
\providecommand \@ifnum [1]{%
 \ifnum #1\expandafter \@firstoftwo
 \else \expandafter \@secondoftwo
 \fi
}%
\providecommand \@ifx [1]{%
 \ifx #1\expandafter \@firstoftwo
 \else \expandafter \@secondoftwo
 \fi
}%
\providecommand \natexlab [1]{#1}%
\providecommand \enquote  [1]{``#1''}%
\providecommand \bibnamefont  [1]{#1}%
\providecommand \bibfnamefont [1]{#1}%
\providecommand \citenamefont [1]{#1}%
\providecommand \href@noop [0]{\@secondoftwo}%
\providecommand \href [0]{\begingroup \@sanitize@url \@href}%
\providecommand \@href[1]{\@@startlink{#1}\@@href}%
\providecommand \@@href[1]{\endgroup#1\@@endlink}%
\providecommand \@sanitize@url [0]{\catcode `\\12\catcode `\$12\catcode
  `\&12\catcode `\#12\catcode `\^12\catcode `\_12\catcode `\%12\relax}%
\providecommand \@@startlink[1]{}%
\providecommand \@@endlink[0]{}%
\providecommand \url  [0]{\begingroup\@sanitize@url \@url }%
\providecommand \@url [1]{\endgroup\@href {#1}{\urlprefix }}%
\providecommand \urlprefix  [0]{URL }%
\providecommand \Eprint [0]{\href }%
\providecommand \doibase [0]{http://dx.doi.org/}%
\providecommand \selectlanguage [0]{\@gobble}%
\providecommand \bibinfo  [0]{\@secondoftwo}%
\providecommand \bibfield  [0]{\@secondoftwo}%
\providecommand \translation [1]{[#1]}%
\providecommand \BibitemOpen [0]{}%
\providecommand \bibitemStop [0]{}%
\providecommand \bibitemNoStop [0]{.\EOS\space}%
\providecommand \EOS [0]{\spacefactor3000\relax}%
\providecommand \BibitemShut  [1]{\csname bibitem#1\endcsname}%
\let\auto@bib@innerbib\@empty
\bibitem [{\citenamefont {Zaanen}(2019)}]{zaanen2019planckian}%
  \BibitemOpen
  \bibfield  {author} {\bibinfo {author} {\bibfnamefont {Jan}\ \bibnamefont
  {Zaanen}},\ }\bibfield  {title} {\enquote {\bibinfo {title} {Planckian
  dissipation, minimal viscosity and the transport in cuprate strange
  metals},}\ }\href@noop {} {\bibfield  {journal} {\bibinfo  {journal} {SciPost
  Physics}\ }\textbf {\bibinfo {volume} {6}},\ \bibinfo {pages} {061} (\bibinfo
  {year} {2019})}\BibitemShut {NoStop}%
\bibitem [{\citenamefont {Ayres}\ \emph {et~al.}(2021)\citenamefont {Ayres},
  \citenamefont {Berben}, \citenamefont {{\v{C}}ulo}, \citenamefont {Hsu},
  \citenamefont {van Heumen}, \citenamefont {Huang}, \citenamefont {Zaanen},
  \citenamefont {Kondo}, \citenamefont {Takeuchi}, \citenamefont {Cooper} \emph
  {et~al.}}]{ayres2021incoherent}%
  \BibitemOpen
  \bibfield  {author} {\bibinfo {author} {\bibfnamefont {J}~\bibnamefont
  {Ayres}}, \bibinfo {author} {\bibfnamefont {M}~\bibnamefont {Berben}},
  \bibinfo {author} {\bibfnamefont {M}~\bibnamefont {{\v{C}}ulo}}, \bibinfo
  {author} {\bibfnamefont {Y-T}\ \bibnamefont {Hsu}}, \bibinfo {author}
  {\bibfnamefont {E}~\bibnamefont {van Heumen}}, \bibinfo {author}
  {\bibfnamefont {Y}~\bibnamefont {Huang}}, \bibinfo {author} {\bibfnamefont
  {J}~\bibnamefont {Zaanen}}, \bibinfo {author} {\bibfnamefont {T}~\bibnamefont
  {Kondo}}, \bibinfo {author} {\bibfnamefont {T}~\bibnamefont {Takeuchi}},
  \bibinfo {author} {\bibfnamefont {JR}~\bibnamefont {Cooper}},  \emph
  {et~al.},\ }\bibfield  {title} {\enquote {\bibinfo {title} {Incoherent
  transport across the strange-metal regime of overdoped cuprates},}\
  }\href@noop {} {\bibfield  {journal} {\bibinfo  {journal} {Nature}\ }\textbf
  {\bibinfo {volume} {595}},\ \bibinfo {pages} {661--666} (\bibinfo {year}
  {2021})}\BibitemShut {NoStop}%
\bibitem [{\citenamefont {Poniatowski}\ \emph {et~al.}(2021)\citenamefont
  {Poniatowski}, \citenamefont {Sarkar}, \citenamefont {Lobo}, \citenamefont
  {Das~Sarma},\ and\ \citenamefont {Greene}}]{poniatowski2021counterexample}%
  \BibitemOpen
  \bibfield  {author} {\bibinfo {author} {\bibfnamefont {Nicholas~R}\
  \bibnamefont {Poniatowski}}, \bibinfo {author} {\bibfnamefont {Tarapada}\
  \bibnamefont {Sarkar}}, \bibinfo {author} {\bibfnamefont {Ricardo~PSM}\
  \bibnamefont {Lobo}}, \bibinfo {author} {\bibfnamefont {Sankar}\ \bibnamefont
  {Das~Sarma}}, \ and\ \bibinfo {author} {\bibfnamefont {Richard~L}\
  \bibnamefont {Greene}},\ }\bibfield  {title} {\enquote {\bibinfo {title}
  {Counterexample to the conjectured planckian bound on transport},}\
  }\href@noop {} {\bibfield  {journal} {\bibinfo  {journal} {Physical Review
  B}\ }\textbf {\bibinfo {volume} {104}},\ \bibinfo {pages} {235138} (\bibinfo
  {year} {2021})}\BibitemShut {NoStop}%
\bibitem [{\citenamefont {Spivak}\ \emph {et~al.}(2010)\citenamefont {Spivak},
  \citenamefont {Kravchenko}, \citenamefont {Kivelson},\ and\ \citenamefont
  {Gao}}]{spivak2010colloquium}%
  \BibitemOpen
  \bibfield  {author} {\bibinfo {author} {\bibfnamefont {B}~\bibnamefont
  {Spivak}}, \bibinfo {author} {\bibfnamefont {SV}~\bibnamefont {Kravchenko}},
  \bibinfo {author} {\bibfnamefont {SA}~\bibnamefont {Kivelson}}, \ and\
  \bibinfo {author} {\bibfnamefont {XPA}\ \bibnamefont {Gao}},\ }\bibfield
  {title} {\enquote {\bibinfo {title} {Colloquium: Transport in strongly
  correlated two dimensional electron fluids},}\ }\href@noop {} {\bibfield
  {journal} {\bibinfo  {journal} {Reviews of modern physics}\ }\textbf
  {\bibinfo {volume} {82}},\ \bibinfo {pages} {1743} (\bibinfo {year}
  {2010})}\BibitemShut {NoStop}%
\bibitem [{\citenamefont {Kasahara}\ \emph {et~al.}(2010)\citenamefont
  {Kasahara}, \citenamefont {Shibauchi}, \citenamefont {Hashimoto},
  \citenamefont {Ikada}, \citenamefont {Tonegawa}, \citenamefont {Okazaki},
  \citenamefont {Shishido}, \citenamefont {Ikeda}, \citenamefont {Takeya},
  \citenamefont {Hirata} \emph {et~al.}}]{kasahara2010evolution}%
  \BibitemOpen
  \bibfield  {author} {\bibinfo {author} {\bibfnamefont {S}~\bibnamefont
  {Kasahara}}, \bibinfo {author} {\bibfnamefont {T}~\bibnamefont {Shibauchi}},
  \bibinfo {author} {\bibfnamefont {K}~\bibnamefont {Hashimoto}}, \bibinfo
  {author} {\bibfnamefont {K}~\bibnamefont {Ikada}}, \bibinfo {author}
  {\bibfnamefont {S}~\bibnamefont {Tonegawa}}, \bibinfo {author} {\bibfnamefont
  {R}~\bibnamefont {Okazaki}}, \bibinfo {author} {\bibfnamefont
  {H}~\bibnamefont {Shishido}}, \bibinfo {author} {\bibfnamefont
  {H}~\bibnamefont {Ikeda}}, \bibinfo {author} {\bibfnamefont {H}~\bibnamefont
  {Takeya}}, \bibinfo {author} {\bibfnamefont {K}~\bibnamefont {Hirata}},
  \emph {et~al.},\ }\bibfield  {title} {\enquote {\bibinfo {title} {Evolution
  from non-fermi-to fermi-liquid transport via isovalent doping in bafe 2 (as
  1- x p x) 2 superconductors},}\ }\href@noop {} {\bibfield  {journal}
  {\bibinfo  {journal} {Physical Review B}\ }\textbf {\bibinfo {volume} {81}},\
  \bibinfo {pages} {184519} (\bibinfo {year} {2010})}\BibitemShut {NoStop}%
\bibitem [{\citenamefont {Sachdev}\ and\ \citenamefont
  {Keimer}(2011)}]{sachdev2011quantum}%
  \BibitemOpen
  \bibfield  {author} {\bibinfo {author} {\bibfnamefont {Subir}\ \bibnamefont
  {Sachdev}}\ and\ \bibinfo {author} {\bibfnamefont {Bernhard}\ \bibnamefont
  {Keimer}},\ }\bibfield  {title} {\enquote {\bibinfo {title} {Quantum
  criticality},}\ }\href@noop {} {\bibfield  {journal} {\bibinfo  {journal}
  {Physics Today}\ }\textbf {\bibinfo {volume} {64}},\ \bibinfo {pages}
  {29--35} (\bibinfo {year} {2011})}\BibitemShut {NoStop}%
\bibitem [{\citenamefont {Lucas}\ and\ \citenamefont
  {Sachdev}(2015)}]{lucasMemoryMatrixTheory2015}%
  \BibitemOpen
  \bibfield  {author} {\bibinfo {author} {\bibfnamefont {Andrew}\ \bibnamefont
  {Lucas}}\ and\ \bibinfo {author} {\bibfnamefont {Subir}\ \bibnamefont
  {Sachdev}},\ }\bibfield  {title} {\enquote {\bibinfo {title} {Memory matrix
  theory of magnetotransport in strange metals},}\ }\href {\doibase
  10.1103/PhysRevB.91.195122} {\bibfield  {journal} {\bibinfo  {journal}
  {Physical Review B}\ }\textbf {\bibinfo {volume} {91}},\ \bibinfo {pages}
  {195122} (\bibinfo {year} {2015})}\BibitemShut {NoStop}%
\bibitem [{\citenamefont {Stephanov}\ \emph {et~al.}(1998)\citenamefont
  {Stephanov}, \citenamefont {Rajagopal},\ and\ \citenamefont
  {Shuryak}}]{stephanov_signatures_1998}%
  \BibitemOpen
  \bibfield  {author} {\bibinfo {author} {\bibfnamefont {M.}~\bibnamefont
  {Stephanov}}, \bibinfo {author} {\bibfnamefont {K.}~\bibnamefont
  {Rajagopal}}, \ and\ \bibinfo {author} {\bibfnamefont {E.}~\bibnamefont
  {Shuryak}},\ }\bibfield  {title} {\enquote {\bibinfo {title} {Signatures of
  the tricritical point in {QCD}},}\ }\href {\doibase
  10.1103/physrevlett.81.4816} {\bibfield  {journal} {\bibinfo  {journal}
  {Physical Review Letters}\ }\textbf {\bibinfo {volume} {81}},\ \bibinfo
  {pages} {4816} (\bibinfo {year} {1998})}\BibitemShut {NoStop}%
\bibitem [{\citenamefont {Kolb}\ \emph {et~al.}(2000)\citenamefont {Kolb},
  \citenamefont {Sollfrank},\ and\ \citenamefont
  {Heinz}}]{kolb_anisotropic_2000}%
  \BibitemOpen
  \bibfield  {author} {\bibinfo {author} {\bibfnamefont {Peter~F.}\
  \bibnamefont {Kolb}}, \bibinfo {author} {\bibfnamefont {Josef}\ \bibnamefont
  {Sollfrank}}, \ and\ \bibinfo {author} {\bibfnamefont {Ulrich}\ \bibnamefont
  {Heinz}},\ }\bibfield  {title} {\enquote {\bibinfo {title} {Anisotropic
  transverse flow and the quark-hadron phase transition},}\ }\href {\doibase
  10.1103/physrevc.62.054909} {\bibfield  {journal} {\bibinfo  {journal}
  {Physical Review C}\ }\textbf {\bibinfo {volume} {62}},\ \bibinfo {pages}
  {054909} (\bibinfo {year} {2000})}\BibitemShut {NoStop}%
\bibitem [{\citenamefont {Ollitrault}(1992)}]{ollitrault_anisotropy_1992}%
  \BibitemOpen
  \bibfield  {author} {\bibinfo {author} {\bibfnamefont {Jean-Yves}\
  \bibnamefont {Ollitrault}},\ }\href {\doibase 10.1103/physrevd.46.229}
  {\bibfield  {journal} {\bibinfo  {journal} {Physical Review D}\ }\textbf
  {\bibinfo {volume} {46}},\ \bibinfo {pages} {229} (\bibinfo {year}
  {1992})}\BibitemShut {NoStop}%
\bibitem [{\citenamefont {{STAR
  Collaboration}}(2010)}]{star_collaboration_experimental_2010}%
  \BibitemOpen
  \bibfield  {author} {\bibinfo {author} {\bibnamefont {{STAR
  Collaboration}}},\ }\bibfield  {title} {\enquote {\bibinfo {title} {An
  {Experimental} {Exploration} of the {QCD} {Phase} {Diagram}: {The} {Search}
  for the {Critical} {Point} and the {Onset} of {De}-confinement},}\ }\href
  {http://arxiv.org/abs/1007.2613} {\bibfield  {journal} {\bibinfo  {journal}
  {arXiv:1007.2613 [nucl-ex]}\ } (\bibinfo {year} {2010})}\BibitemShut
  {NoStop}%
\bibitem [{\citenamefont {Heinz}\ \emph {et~al.}(2015)\citenamefont {Heinz},
  \citenamefont {Sorensen}, \citenamefont {Deshpande}, \citenamefont
  {Gagliardi}, \citenamefont {Karsch}, \citenamefont {Lappi}, \citenamefont
  {Meziani}, \citenamefont {Milner}, \citenamefont {Muller}, \citenamefont
  {Nagle}, \citenamefont {Qiu}, \citenamefont {Rajagopal}, \citenamefont
  {Roland},\ and\ \citenamefont {Venugopalan}}]{heinz_exploring_2015}%
  \BibitemOpen
  \bibfield  {author} {\bibinfo {author} {\bibfnamefont {U.}~\bibnamefont
  {Heinz}}, \bibinfo {author} {\bibfnamefont {P.}~\bibnamefont {Sorensen}},
  \bibinfo {author} {\bibfnamefont {A.}~\bibnamefont {Deshpande}}, \bibinfo
  {author} {\bibfnamefont {C.}~\bibnamefont {Gagliardi}}, \bibinfo {author}
  {\bibfnamefont {F.}~\bibnamefont {Karsch}}, \bibinfo {author} {\bibfnamefont
  {T.}~\bibnamefont {Lappi}}, \bibinfo {author} {\bibfnamefont {Z.-E.}\
  \bibnamefont {Meziani}}, \bibinfo {author} {\bibfnamefont {R.}~\bibnamefont
  {Milner}}, \bibinfo {author} {\bibfnamefont {B.}~\bibnamefont {Muller}},
  \bibinfo {author} {\bibfnamefont {J.}~\bibnamefont {Nagle}}, \bibinfo
  {author} {\bibfnamefont {J.-W.}\ \bibnamefont {Qiu}}, \bibinfo {author}
  {\bibfnamefont {K.}~\bibnamefont {Rajagopal}}, \bibinfo {author}
  {\bibfnamefont {G.}~\bibnamefont {Roland}}, \ and\ \bibinfo {author}
  {\bibfnamefont {R.}~\bibnamefont {Venugopalan}},\ }\bibfield  {title}
  {\enquote {\bibinfo {title} {Exploring the properties of the phases of {QCD}
  matter - research opportunities and priorities for the next decade},}\ }\href
  {http://arxiv.org/abs/1501.06477} {\bibfield  {journal} {\bibinfo  {journal}
  {arXiv:1501.06477 [hep-ex, physics:hep-ph, physics:nucl-ex,
  physics:nucl-th]}\ } (\bibinfo {year} {2015})}\BibitemShut {NoStop}%
\bibitem [{\citenamefont {Pitaevskii}\ and\ \citenamefont
  {Lifshitz}(1981)}]{pitaevskiiPhysicalKineticsVolume1981}%
  \BibitemOpen
  \bibfield  {author} {\bibinfo {author} {\bibfnamefont {L.~P.}\ \bibnamefont
  {Pitaevskii}}\ and\ \bibinfo {author} {\bibfnamefont {E.~M.}\ \bibnamefont
  {Lifshitz}},\ }\href@noop {} {\emph {\bibinfo {title} {Physical {Kinetics}:
  {Volume} 10}}},\ \bibinfo {edition} {1st}\ ed.\ (\bibinfo  {publisher}
  {Butterworth-Heinemann},\ \bibinfo {address} {Amsterdam},\ \bibinfo {year}
  {1981})\BibitemShut {NoStop}%
\bibitem [{\citenamefont {Stark}\ and\ \citenamefont
  {Kollar}(2013)}]{starkKineticDescriptionThermalization2013}%
  \BibitemOpen
  \bibfield  {author} {\bibinfo {author} {\bibfnamefont {Michael}\ \bibnamefont
  {Stark}}\ and\ \bibinfo {author} {\bibfnamefont {Marcus}\ \bibnamefont
  {Kollar}},\ }\href {\doibase 10.48550/arXiv.1308.1610} {\enquote {\bibinfo
  {title} {Kinetic description of thermalization dynamics in weakly interacting
  quantum systems},}\ } (\bibinfo {year} {2013}),\ \Eprint
  {http://arxiv.org/abs/1308.1610} {arxiv:1308.1610 [cond-mat]} \BibitemShut
  {NoStop}%
\bibitem [{\citenamefont {Bertini}\ \emph {et~al.}(2015)\citenamefont
  {Bertini}, \citenamefont {Essler}, \citenamefont {Groha},\ and\ \citenamefont
  {Robinson}}]{bertiniPrethermalizationThermalizationModels2015}%
  \BibitemOpen
  \bibfield  {author} {\bibinfo {author} {\bibfnamefont {Bruno}\ \bibnamefont
  {Bertini}}, \bibinfo {author} {\bibfnamefont {Fabian H.~L.}\ \bibnamefont
  {Essler}}, \bibinfo {author} {\bibfnamefont {Stefan}\ \bibnamefont {Groha}},
  \ and\ \bibinfo {author} {\bibfnamefont {Neil~J.}\ \bibnamefont {Robinson}},\
  }\bibfield  {title} {\enquote {\bibinfo {title} {Prethermalization and
  {{Thermalization}} in {{Models}} with {{Weak Integrability Breaking}}},}\
  }\href {\doibase 10.1103/PhysRevLett.115.180601} {\bibfield  {journal}
  {\bibinfo  {journal} {Physical Review Letters}\ }\textbf {\bibinfo {volume}
  {115}},\ \bibinfo {pages} {180601} (\bibinfo {year} {2015})}\BibitemShut
  {NoStop}%
\bibitem [{\citenamefont {Mallayya}\ \emph {et~al.}(2019)\citenamefont
  {Mallayya}, \citenamefont {Rigol},\ and\ \citenamefont
  {De~Roeck}}]{mallayyaPrethermalizationThermalizationIsolated2019}%
  \BibitemOpen
  \bibfield  {author} {\bibinfo {author} {\bibfnamefont {Krishnanand}\
  \bibnamefont {Mallayya}}, \bibinfo {author} {\bibfnamefont {Marcos}\
  \bibnamefont {Rigol}}, \ and\ \bibinfo {author} {\bibfnamefont {Wojciech}\
  \bibnamefont {De~Roeck}},\ }\bibfield  {title} {\enquote {\bibinfo {title}
  {Prethermalization and {{Thermalization}} in {{Isolated Quantum Systems}}},}\
  }\href {\doibase 10.1103/PhysRevX.9.021027} {\bibfield  {journal} {\bibinfo
  {journal} {Physical Review X}\ }\textbf {\bibinfo {volume} {9}},\ \bibinfo
  {pages} {021027} (\bibinfo {year} {2019})}\BibitemShut {NoStop}%
\bibitem [{\citenamefont {Friedman}\ \emph {et~al.}(2020)\citenamefont
  {Friedman}, \citenamefont {Gopalakrishnan},\ and\ \citenamefont
  {Vasseur}}]{friedmanDiffusiveHydrodynamicsIntegrability2020}%
  \BibitemOpen
  \bibfield  {author} {\bibinfo {author} {\bibfnamefont {Aaron~J.}\
  \bibnamefont {Friedman}}, \bibinfo {author} {\bibfnamefont {Sarang}\
  \bibnamefont {Gopalakrishnan}}, \ and\ \bibinfo {author} {\bibfnamefont
  {Romain}\ \bibnamefont {Vasseur}},\ }\bibfield  {title} {\enquote {\bibinfo
  {title} {Diffusive hydrodynamics from integrability breaking},}\ }\href
  {\doibase 10.1103/PhysRevB.101.180302} {\bibfield  {journal} {\bibinfo
  {journal} {Physical Review B}\ }\textbf {\bibinfo {volume} {101}},\ \bibinfo
  {pages} {180302} (\bibinfo {year} {2020})},\ \bibinfo {note} {publisher:
  American Physical Society}\BibitemShut {NoStop}%
\bibitem [{\citenamefont {Durnin}\ \emph {et~al.}(2021)\citenamefont {Durnin},
  \citenamefont {Bhaseen},\ and\ \citenamefont
  {Doyon}}]{durninNonEquilibriumDynamicsWeakly2021}%
  \BibitemOpen
  \bibfield  {author} {\bibinfo {author} {\bibfnamefont {Joseph}\ \bibnamefont
  {Durnin}}, \bibinfo {author} {\bibfnamefont {M.~J.}\ \bibnamefont {Bhaseen}},
  \ and\ \bibinfo {author} {\bibfnamefont {Benjamin}\ \bibnamefont {Doyon}},\
  }\bibfield  {title} {\enquote {\bibinfo {title} {Non-{Equilibrium} {Dynamics}
  and {Weakly} {Broken} {Integrability}},}\ }\href {\doibase
  10.1103/PhysRevLett.127.130601} {\bibfield  {journal} {\bibinfo  {journal}
  {Physical Review Letters}\ }\textbf {\bibinfo {volume} {127}},\ \bibinfo
  {pages} {130601} (\bibinfo {year} {2021})},\ \bibinfo {note}
  {arXiv:2004.11030 [cond-mat, physics:hep-th, physics:math-ph]}\BibitemShut
  {NoStop}%
\bibitem [{\citenamefont {Lopez-Piqueres}\ \emph {et~al.}(2021)\citenamefont
  {Lopez-Piqueres}, \citenamefont {Ware}, \citenamefont {Gopalakrishnan},\ and\
  \citenamefont
  {Vasseur}}]{lopez-piqueresHydrodynamicsNonintegrableSystems2021}%
  \BibitemOpen
  \bibfield  {author} {\bibinfo {author} {\bibfnamefont {Javier}\ \bibnamefont
  {Lopez-Piqueres}}, \bibinfo {author} {\bibfnamefont {Brayden}\ \bibnamefont
  {Ware}}, \bibinfo {author} {\bibfnamefont {Sarang}\ \bibnamefont
  {Gopalakrishnan}}, \ and\ \bibinfo {author} {\bibfnamefont {Romain}\
  \bibnamefont {Vasseur}},\ }\bibfield  {title} {\enquote {\bibinfo {title}
  {Hydrodynamics of nonintegrable systems from a relaxation-time
  approximation},}\ }\href {\doibase 10.1103/PhysRevB.103.L060302} {\bibfield
  {journal} {\bibinfo  {journal} {Physical Review B}\ }\textbf {\bibinfo
  {volume} {103}},\ \bibinfo {pages} {L060302} (\bibinfo {year} {2021})},\
  \bibinfo {note} {arXiv: 2005.13546}\BibitemShut {NoStop}%
\bibitem [{\citenamefont {De~Nardis}\ \emph {et~al.}(2021)\citenamefont
  {De~Nardis}, \citenamefont {Gopalakrishnan}, \citenamefont {Vasseur},\ and\
  \citenamefont {Ware}}]{denardisStabilitySuperdiffusionNearly2021a}%
  \BibitemOpen
  \bibfield  {author} {\bibinfo {author} {\bibfnamefont {Jacopo}\ \bibnamefont
  {De~Nardis}}, \bibinfo {author} {\bibfnamefont {Sarang}\ \bibnamefont
  {Gopalakrishnan}}, \bibinfo {author} {\bibfnamefont {Romain}\ \bibnamefont
  {Vasseur}}, \ and\ \bibinfo {author} {\bibfnamefont {Brayden}\ \bibnamefont
  {Ware}},\ }\bibfield  {title} {\enquote {\bibinfo {title} {Stability of
  superdiffusion in nearly integrable spin chains},}\ }\href {\doibase
  10.1103/PhysRevLett.127.057201} {\bibfield  {journal} {\bibinfo  {journal}
  {Physical Review Letters}\ }\textbf {\bibinfo {volume} {127}},\ \bibinfo
  {pages} {057201} (\bibinfo {year} {2021})},\ \Eprint
  {http://arxiv.org/abs/2102.02219} {arxiv:2102.02219} \BibitemShut {NoStop}%
\bibitem [{\citenamefont {Khemani}\ \emph {et~al.}(2018)\citenamefont
  {Khemani}, \citenamefont {Vishwanath},\ and\ \citenamefont
  {Huse}}]{khemaniOperatorSpreadingEmergence2018}%
  \BibitemOpen
  \bibfield  {author} {\bibinfo {author} {\bibfnamefont {Vedika}\ \bibnamefont
  {Khemani}}, \bibinfo {author} {\bibfnamefont {Ashvin}\ \bibnamefont
  {Vishwanath}}, \ and\ \bibinfo {author} {\bibfnamefont {David~A.}\
  \bibnamefont {Huse}},\ }\bibfield  {title} {\enquote {\bibinfo {title}
  {Operator {{Spreading}} and the {{Emergence}} of {{Dissipative
  Hydrodynamics}} under {{Unitary Evolution}} with {{Conservation Laws}}},}\
  }\href {\doibase 10.1103/PhysRevX.8.031057} {\bibfield  {journal} {\bibinfo
  {journal} {Physical Review X}\ }\textbf {\bibinfo {volume} {8}},\ \bibinfo
  {pages} {031057} (\bibinfo {year} {2018})}\BibitemShut {NoStop}%
\bibitem [{\citenamefont {Kvorning}\ \emph {et~al.}(2021)\citenamefont
  {Kvorning}, \citenamefont {Herviou},\ and\ \citenamefont
  {Bardarson}}]{kvorningTimeevolutionLocalInformation2021}%
  \BibitemOpen
  \bibfield  {author} {\bibinfo {author} {\bibfnamefont {Thomas~Klein}\
  \bibnamefont {Kvorning}}, \bibinfo {author} {\bibfnamefont {Lo{\"i}c}\
  \bibnamefont {Herviou}}, \ and\ \bibinfo {author} {\bibfnamefont {Jens~H.}\
  \bibnamefont {Bardarson}},\ }\bibfield  {title} {\enquote {\bibinfo {title}
  {Time-evolution of local information: Thermalization dynamics of local
  observables},}\ }\href {http://arxiv.org/abs/2105.11206} {\bibfield
  {journal} {\bibinfo  {journal} {arXiv:2105.11206 [cond-mat,
  physics:quant-ph]}\ } (\bibinfo {year} {2021})},\ \Eprint
  {http://arxiv.org/abs/2105.11206} {arxiv:2105.11206 [cond-mat,
  physics:quant-ph]} \BibitemShut {NoStop}%
\bibitem [{\citenamefont {{von Keyserlingk}}\ \emph {et~al.}(2022)\citenamefont
  {{von Keyserlingk}}, \citenamefont {Pollmann},\ and\ \citenamefont
  {Rakovszky}}]{vonkeyserlingkOperatorBackflowClassical2022}%
  \BibitemOpen
  \bibfield  {author} {\bibinfo {author} {\bibfnamefont {Curt}\ \bibnamefont
  {{von Keyserlingk}}}, \bibinfo {author} {\bibfnamefont {Frank}\ \bibnamefont
  {Pollmann}}, \ and\ \bibinfo {author} {\bibfnamefont {Tibor}\ \bibnamefont
  {Rakovszky}},\ }\bibfield  {title} {\enquote {\bibinfo {title} {Operator
  backflow and the classical simulation of quantum transport},}\ }\href
  {\doibase 10.1103/PhysRevB.105.245101} {\bibfield  {journal} {\bibinfo
  {journal} {Physical Review B}\ }\textbf {\bibinfo {volume} {105}},\ \bibinfo
  {pages} {245101} (\bibinfo {year} {2022})}\BibitemShut {NoStop}%
\bibitem [{\citenamefont {White}(2023)}]{whiteEffectiveDissipationRate2023}%
  \BibitemOpen
  \bibfield  {author} {\bibinfo {author} {\bibfnamefont {Christopher~David}\
  \bibnamefont {White}},\ }\bibfield  {title} {\enquote {\bibinfo {title}
  {Effective dissipation rate in a {{Liouvillian-graph}} picture of
  high-temperature quantum hydrodynamics},}\ }\href {\doibase
  10.1103/PhysRevB.107.094311} {\bibfield  {journal} {\bibinfo  {journal}
  {Physical Review B}\ }\textbf {\bibinfo {volume} {107}},\ \bibinfo {pages}
  {094311} (\bibinfo {year} {2023})}\BibitemShut {NoStop}%
\bibitem [{\citenamefont {Artiaco}\ \emph {et~al.}(2023)\citenamefont
  {Artiaco}, \citenamefont {Fleckenstein}, \citenamefont {Aceituno},
  \citenamefont {Kvorning},\ and\ \citenamefont
  {Bardarson}}]{artiacoEfficientLargeScaleManyBody2023}%
  \BibitemOpen
  \bibfield  {author} {\bibinfo {author} {\bibfnamefont {Claudia}\ \bibnamefont
  {Artiaco}}, \bibinfo {author} {\bibfnamefont {Christoph}\ \bibnamefont
  {Fleckenstein}}, \bibinfo {author} {\bibfnamefont {David}\ \bibnamefont
  {Aceituno}}, \bibinfo {author} {\bibfnamefont {Thomas~Klein}\ \bibnamefont
  {Kvorning}}, \ and\ \bibinfo {author} {\bibfnamefont {Jens~H.}\ \bibnamefont
  {Bardarson}},\ }\href {\doibase 10.48550/arXiv.2310.06036} {\enquote
  {\bibinfo {title} {Efficient {{Large-Scale Many-Body Quantum Dynamics}} via
  {{Local-Information Time Evolution}}},}\ } (\bibinfo {year} {2023}),\ \Eprint
  {http://arxiv.org/abs/2310.06036} {arxiv:2310.06036 [cond-mat,
  physics:quant-ph]} \BibitemShut {NoStop}%
\bibitem [{\citenamefont {Brown}\ \emph {et~al.}(2019)\citenamefont {Brown},
  \citenamefont {Mitra}, \citenamefont {Guardado-Sanchez}, \citenamefont
  {Nourafkan}, \citenamefont {Reymbaut}, \citenamefont {Hébert}, \citenamefont
  {Bergeron}, \citenamefont {Tremblay}, \citenamefont {Kokalj}, \citenamefont
  {Huse}, \citenamefont {Schauß},\ and\ \citenamefont
  {Bakr}}]{brownBadMetallicTransport2019a}%
  \BibitemOpen
  \bibfield  {author} {\bibinfo {author} {\bibfnamefont {Peter~T.}\
  \bibnamefont {Brown}}, \bibinfo {author} {\bibfnamefont {Debayan}\
  \bibnamefont {Mitra}}, \bibinfo {author} {\bibfnamefont {Elmer}\ \bibnamefont
  {Guardado-Sanchez}}, \bibinfo {author} {\bibfnamefont {Reza}\ \bibnamefont
  {Nourafkan}}, \bibinfo {author} {\bibfnamefont {Alexis}\ \bibnamefont
  {Reymbaut}}, \bibinfo {author} {\bibfnamefont {Charles-David}\ \bibnamefont
  {Hébert}}, \bibinfo {author} {\bibfnamefont {Simon}\ \bibnamefont
  {Bergeron}}, \bibinfo {author} {\bibfnamefont {A.-M.~S.}\ \bibnamefont
  {Tremblay}}, \bibinfo {author} {\bibfnamefont {Jure}\ \bibnamefont {Kokalj}},
  \bibinfo {author} {\bibfnamefont {David~A.}\ \bibnamefont {Huse}}, \bibinfo
  {author} {\bibfnamefont {Peter}\ \bibnamefont {Schauß}}, \ and\ \bibinfo
  {author} {\bibfnamefont {Waseem~S.}\ \bibnamefont {Bakr}},\ }\bibfield
  {title} {\enquote {\bibinfo {title} {Bad metallic transport in a cold atom
  {Fermi}-{Hubbard} system},}\ }\href {\doibase 10.1126/science.aat4134}
  {\bibfield  {journal} {\bibinfo  {journal} {Science}\ }\textbf {\bibinfo
  {volume} {363}},\ \bibinfo {pages} {379--382} (\bibinfo {year} {2019})},\
  \bibinfo {note} {publisher: American Association for the Advancement of
  Science}\BibitemShut {NoStop}%
\bibitem [{\citenamefont {Yan}\ \emph {et~al.}(2022)\citenamefont {Yan},
  \citenamefont {Spar}, \citenamefont {Prichard}, \citenamefont {Chi},
  \citenamefont {Wei}, \citenamefont {Ibarra-García-Padilla}, \citenamefont
  {Hazzard},\ and\ \citenamefont
  {Bakr}}]{yanTwodimensionalProgrammableTweezer2022}%
  \BibitemOpen
  \bibfield  {author} {\bibinfo {author} {\bibfnamefont {Zoe~Z.}\ \bibnamefont
  {Yan}}, \bibinfo {author} {\bibfnamefont {Benjamin~M.}\ \bibnamefont {Spar}},
  \bibinfo {author} {\bibfnamefont {Max~L.}\ \bibnamefont {Prichard}}, \bibinfo
  {author} {\bibfnamefont {Sungjae}\ \bibnamefont {Chi}}, \bibinfo {author}
  {\bibfnamefont {Hao-Tian}\ \bibnamefont {Wei}}, \bibinfo {author}
  {\bibfnamefont {Eduardo}\ \bibnamefont {Ibarra-García-Padilla}}, \bibinfo
  {author} {\bibfnamefont {Kaden R.~A.}\ \bibnamefont {Hazzard}}, \ and\
  \bibinfo {author} {\bibfnamefont {Waseem~S.}\ \bibnamefont {Bakr}},\
  }\bibfield  {title} {\enquote {\bibinfo {title} {A two-dimensional
  programmable tweezer array of fermions},}\ }\href {\doibase
  10.1103/PhysRevLett.129.123201} {\bibfield  {journal} {\bibinfo  {journal}
  {Physical Review Letters}\ }\textbf {\bibinfo {volume} {129}},\ \bibinfo
  {pages} {123201} (\bibinfo {year} {2022})},\ \bibinfo {note}
  {arXiv:2203.15023 [cond-mat, physics:physics, physics:quant-ph]}\BibitemShut
  {NoStop}%
\bibitem [{\citenamefont {Wienand}\ \emph {et~al.}(2023)\citenamefont
  {Wienand}, \citenamefont {Karch}, \citenamefont {Impertro}, \citenamefont
  {Schweizer}, \citenamefont {McCulloch}, \citenamefont {Vasseur},
  \citenamefont {Gopalakrishnan}, \citenamefont {Aidelsburger},\ and\
  \citenamefont {Bloch}}]{wienandEmergenceFluctuatingHydrodynamics2023}%
  \BibitemOpen
  \bibfield  {author} {\bibinfo {author} {\bibfnamefont {Julian~F.}\
  \bibnamefont {Wienand}}, \bibinfo {author} {\bibfnamefont {Simon}\
  \bibnamefont {Karch}}, \bibinfo {author} {\bibfnamefont {Alexander}\
  \bibnamefont {Impertro}}, \bibinfo {author} {\bibfnamefont {Christian}\
  \bibnamefont {Schweizer}}, \bibinfo {author} {\bibfnamefont {Ewan}\
  \bibnamefont {McCulloch}}, \bibinfo {author} {\bibfnamefont {Romain}\
  \bibnamefont {Vasseur}}, \bibinfo {author} {\bibfnamefont {Sarang}\
  \bibnamefont {Gopalakrishnan}}, \bibinfo {author} {\bibfnamefont {Monika}\
  \bibnamefont {Aidelsburger}}, \ and\ \bibinfo {author} {\bibfnamefont
  {Immanuel}\ \bibnamefont {Bloch}},\ }\href {http://arxiv.org/abs/2306.11457}
  {\enquote {\bibinfo {title} {Emergence of fluctuating hydrodynamics in
  chaotic quantum systems},}\ } (\bibinfo {year} {2023}),\ \bibinfo {note}
  {arXiv:2306.11457 [cond-mat, physics:quant-ph]}\BibitemShut {NoStop}%
\bibitem [{\citenamefont {Bertini}\ \emph {et~al.}(2021)\citenamefont
  {Bertini}, \citenamefont {Heidrich-Meisner}, \citenamefont {Karrasch},
  \citenamefont {Prosen}, \citenamefont {Steinigeweg},\ and\ \citenamefont
  {Znidaric}}]{bertiniFinitetemperatureTransportOnedimensional2021a}%
  \BibitemOpen
  \bibfield  {author} {\bibinfo {author} {\bibfnamefont {B.}~\bibnamefont
  {Bertini}}, \bibinfo {author} {\bibfnamefont {F.}~\bibnamefont
  {Heidrich-Meisner}}, \bibinfo {author} {\bibfnamefont {C.}~\bibnamefont
  {Karrasch}}, \bibinfo {author} {\bibfnamefont {T.}~\bibnamefont {Prosen}},
  \bibinfo {author} {\bibfnamefont {R.}~\bibnamefont {Steinigeweg}}, \ and\
  \bibinfo {author} {\bibfnamefont {M.}~\bibnamefont {Znidaric}},\ }\bibfield
  {title} {\enquote {\bibinfo {title} {Finite-temperature transport in
  one-dimensional quantum lattice models},}\ }\href {\doibase
  10.1103/RevModPhys.93.025003} {\bibfield  {journal} {\bibinfo  {journal}
  {Reviews of Modern Physics}\ }\textbf {\bibinfo {volume} {93}},\ \bibinfo
  {pages} {025003} (\bibinfo {year} {2021})},\ \bibinfo {note} {arXiv:
  2003.03334}\BibitemShut {NoStop}%
\bibitem [{\citenamefont {White}\ \emph {et~al.}(2017)\citenamefont {White},
  \citenamefont {Zaletel}, \citenamefont {Mong},\ and\ \citenamefont
  {Refael}}]{whiteQuantumDynamicsThermalizing2017}%
  \BibitemOpen
  \bibfield  {author} {\bibinfo {author} {\bibfnamefont {Christopher~David}\
  \bibnamefont {White}}, \bibinfo {author} {\bibfnamefont {Michael}\
  \bibnamefont {Zaletel}}, \bibinfo {author} {\bibfnamefont {Roger S.~K.}\
  \bibnamefont {Mong}}, \ and\ \bibinfo {author} {\bibfnamefont {Gil}\
  \bibnamefont {Refael}},\ }\bibfield  {title} {\enquote {\bibinfo {title}
  {Quantum dynamics of thermalizing systems},}\ }\href {\doibase
  10.1103/PhysRevB.97.035127} {\  (\bibinfo {year} {2017}),\
  10.1103/PhysRevB.97.035127}\BibitemShut {NoStop}%
\bibitem [{\citenamefont {Rakovszky}\ \emph {et~al.}(2022)\citenamefont
  {Rakovszky}, \citenamefont {von Keyserlingk},\ and\ \citenamefont
  {Pollmann}}]{rakovszky2022dissipation}%
  \BibitemOpen
  \bibfield  {author} {\bibinfo {author} {\bibfnamefont {Tibor}\ \bibnamefont
  {Rakovszky}}, \bibinfo {author} {\bibfnamefont {CW}~\bibnamefont {von
  Keyserlingk}}, \ and\ \bibinfo {author} {\bibfnamefont {Frank}\ \bibnamefont
  {Pollmann}},\ }\bibfield  {title} {\enquote {\bibinfo {title}
  {Dissipation-assisted operator evolution method for capturing hydrodynamic
  transport},}\ }\href@noop {} {\bibfield  {journal} {\bibinfo  {journal}
  {Physical Review B}\ }\textbf {\bibinfo {volume} {105}},\ \bibinfo {pages}
  {075131} (\bibinfo {year} {2022})}\BibitemShut {NoStop}%
\bibitem [{\citenamefont {Ye}\ \emph {et~al.}(2020)\citenamefont {Ye},
  \citenamefont {Machado}, \citenamefont {White}, \citenamefont {Mong},\ and\
  \citenamefont {Yao}}]{yeEmergentHydrodynamicsNonequilibrium2020}%
  \BibitemOpen
  \bibfield  {author} {\bibinfo {author} {\bibfnamefont {Bingtian}\
  \bibnamefont {Ye}}, \bibinfo {author} {\bibfnamefont {Francisco}\
  \bibnamefont {Machado}}, \bibinfo {author} {\bibfnamefont
  {Christopher~David}\ \bibnamefont {White}}, \bibinfo {author} {\bibfnamefont
  {Roger S.~K.}\ \bibnamefont {Mong}}, \ and\ \bibinfo {author} {\bibfnamefont
  {Norman~Y.}\ \bibnamefont {Yao}},\ }\bibfield  {title} {\enquote {\bibinfo
  {title} {Emergent hydrodynamics in non-equilibrium quantum systems},}\ }\href
  {\doibase 10.1103/PhysRevLett.125.030601} {\bibfield  {journal} {\bibinfo
  {journal} {Physical Review Letters}\ }\textbf {\bibinfo {volume} {125}},\
  \bibinfo {pages} {030601} (\bibinfo {year} {2020})},\ \Eprint
  {http://arxiv.org/abs/1902.01859} {arxiv:1902.01859} \BibitemShut {NoStop}%
\bibitem [{\citenamefont {Wei}\ \emph {et~al.}(2022)\citenamefont {Wei},
  \citenamefont {{Rubio-Abadal}}, \citenamefont {Ye}, \citenamefont {Machado},
  \citenamefont {Kemp}, \citenamefont {Srakaew}, \citenamefont {Hollerith},
  \citenamefont {Rui}, \citenamefont {Gopalakrishnan}, \citenamefont {Yao},
  \citenamefont {Bloch},\ and\ \citenamefont
  {Zeiher}}]{weiQuantumGasMicroscopy2022}%
  \BibitemOpen
  \bibfield  {author} {\bibinfo {author} {\bibfnamefont {David}\ \bibnamefont
  {Wei}}, \bibinfo {author} {\bibfnamefont {Antonio}\ \bibnamefont
  {{Rubio-Abadal}}}, \bibinfo {author} {\bibfnamefont {Bingtian}\ \bibnamefont
  {Ye}}, \bibinfo {author} {\bibfnamefont {Francisco}\ \bibnamefont {Machado}},
  \bibinfo {author} {\bibfnamefont {Jack}\ \bibnamefont {Kemp}}, \bibinfo
  {author} {\bibfnamefont {Kritsana}\ \bibnamefont {Srakaew}}, \bibinfo
  {author} {\bibfnamefont {Simon}\ \bibnamefont {Hollerith}}, \bibinfo {author}
  {\bibfnamefont {Jun}\ \bibnamefont {Rui}}, \bibinfo {author} {\bibfnamefont
  {Sarang}\ \bibnamefont {Gopalakrishnan}}, \bibinfo {author} {\bibfnamefont
  {Norman~Y.}\ \bibnamefont {Yao}}, \bibinfo {author} {\bibfnamefont
  {Immanuel}\ \bibnamefont {Bloch}}, \ and\ \bibinfo {author} {\bibfnamefont
  {Johannes}\ \bibnamefont {Zeiher}},\ }\bibfield  {title} {\enquote {\bibinfo
  {title} {Quantum gas microscopy of {{Kardar-Parisi-Zhang}} superdiffusion},}\
  }\href {\doibase 10.1126/science.abk2397} {\bibfield  {journal} {\bibinfo
  {journal} {Science}\ }\textbf {\bibinfo {volume} {376}},\ \bibinfo {pages}
  {716--720} (\bibinfo {year} {2022})},\ \Eprint
  {http://arxiv.org/abs/2107.00038} {arxiv:2107.00038 [cond-mat,
  physics:quant-ph]} \BibitemShut {NoStop}%
\bibitem [{\citenamefont {Ye}\ \emph {et~al.}(2022)\citenamefont {Ye},
  \citenamefont {Machado}, \citenamefont {Kemp}, \citenamefont {Hutson},\ and\
  \citenamefont {Yao}}]{yeUniversalKardarParisiZhangDynamics2022b}%
  \BibitemOpen
  \bibfield  {author} {\bibinfo {author} {\bibfnamefont {Bingtian}\
  \bibnamefont {Ye}}, \bibinfo {author} {\bibfnamefont {Francisco}\
  \bibnamefont {Machado}}, \bibinfo {author} {\bibfnamefont {Jack}\
  \bibnamefont {Kemp}}, \bibinfo {author} {\bibfnamefont {Ross~B.}\
  \bibnamefont {Hutson}}, \ and\ \bibinfo {author} {\bibfnamefont {Norman~Y.}\
  \bibnamefont {Yao}},\ }\bibfield  {title} {\enquote {\bibinfo {title}
  {Universal {{Kardar-Parisi-Zhang Dynamics}} in {{Integrable Quantum
  Systems}}},}\ }\href {\doibase 10.1103/PhysRevLett.129.230602} {\bibfield
  {journal} {\bibinfo  {journal} {Physical Review Letters}\ }\textbf {\bibinfo
  {volume} {129}},\ \bibinfo {pages} {230602} (\bibinfo {year}
  {2022})}\BibitemShut {NoStop}%
\bibitem [{\citenamefont {Thomas}\ \emph {et~al.}(2023)\citenamefont {Thomas},
  \citenamefont {Ware}, \citenamefont {Sau},\ and\ \citenamefont
  {White}}]{thomasComparingNumericalMethods2023}%
  \BibitemOpen
  \bibfield  {author} {\bibinfo {author} {\bibfnamefont {Stuart~N.}\
  \bibnamefont {Thomas}}, \bibinfo {author} {\bibfnamefont {Brayden}\
  \bibnamefont {Ware}}, \bibinfo {author} {\bibfnamefont {Jay~D.}\ \bibnamefont
  {Sau}}, \ and\ \bibinfo {author} {\bibfnamefont {Christopher~David}\
  \bibnamefont {White}},\ }\href {\doibase 10.48550/arXiv.2310.06886} {\enquote
  {\bibinfo {title} {Comparing numerical methods for hydrodynamics in a {{1D}}
  lattice model},}\ } (\bibinfo {year} {2023}),\ \Eprint
  {http://arxiv.org/abs/2310.06886} {arxiv:2310.06886 [cond-mat]} \BibitemShut
  {NoStop}%
\bibitem [{\citenamefont {Li}\ and\ \citenamefont
  {Benjamin}(2017)}]{liEfficientVariationalQuantum2017}%
  \BibitemOpen
  \bibfield  {author} {\bibinfo {author} {\bibfnamefont {Ying}\ \bibnamefont
  {Li}}\ and\ \bibinfo {author} {\bibfnamefont {Simon~C.}\ \bibnamefont
  {Benjamin}},\ }\bibfield  {title} {\enquote {\bibinfo {title} {Efficient
  {Variational} {Quantum} {Simulator} {Incorporating} {Active} {Error}
  {Minimization}},}\ }\href {\doibase 10.1103/PhysRevX.7.021050} {\bibfield
  {journal} {\bibinfo  {journal} {Physical Review X}\ }\textbf {\bibinfo
  {volume} {7}},\ \bibinfo {pages} {021050} (\bibinfo {year} {2017})},\
  \bibinfo {note} {publisher: American Physical Society}\BibitemShut {NoStop}%
\bibitem [{\citenamefont {Temme}\ \emph {et~al.}(2017)\citenamefont {Temme},
  \citenamefont {Bravyi},\ and\ \citenamefont
  {Gambetta}}]{temmeErrorMitigationShortDepth2017a}%
  \BibitemOpen
  \bibfield  {author} {\bibinfo {author} {\bibfnamefont {Kristan}\ \bibnamefont
  {Temme}}, \bibinfo {author} {\bibfnamefont {Sergey}\ \bibnamefont {Bravyi}},
  \ and\ \bibinfo {author} {\bibfnamefont {Jay~M.}\ \bibnamefont {Gambetta}},\
  }\bibfield  {title} {\enquote {\bibinfo {title} {Error {Mitigation} for
  {Short}-{Depth} {Quantum} {Circuits}},}\ }\href {\doibase
  10.1103/PhysRevLett.119.180509} {\bibfield  {journal} {\bibinfo  {journal}
  {Physical Review Letters}\ }\textbf {\bibinfo {volume} {119}},\ \bibinfo
  {pages} {180509} (\bibinfo {year} {2017})},\ \bibinfo {note} {publisher:
  American Physical Society}\BibitemShut {NoStop}%
\bibitem [{\citenamefont {Surace}\ and\ \citenamefont
  {Motrunich}(2023{\natexlab{a}})}]{suraceWeakIntegrabilityBreaking2023}%
  \BibitemOpen
  \bibfield  {author} {\bibinfo {author} {\bibfnamefont {Federica~Maria}\
  \bibnamefont {Surace}}\ and\ \bibinfo {author} {\bibfnamefont {Olexei}\
  \bibnamefont {Motrunich}},\ }\href {http://arxiv.org/abs/2302.12804}
  {\enquote {\bibinfo {title} {Weak integrability breaking perturbations of
  integrable models},}\ } (\bibinfo {year} {2023}{\natexlab{a}}),\ \Eprint
  {http://arxiv.org/abs/2302.12804} {arxiv:2302.12804 [cond-mat,
  physics:quant-ph]} \BibitemShut {NoStop}%
\bibitem [{\citenamefont {Rahmani}\ \emph {et~al.}(2015)\citenamefont
  {Rahmani}, \citenamefont {Zhu}, \citenamefont {Franz},\ and\ \citenamefont
  {Affleck}}]{rahmani2015phase}%
  \BibitemOpen
  \bibfield  {author} {\bibinfo {author} {\bibfnamefont {Armin}\ \bibnamefont
  {Rahmani}}, \bibinfo {author} {\bibfnamefont {Xiaoyu}\ \bibnamefont {Zhu}},
  \bibinfo {author} {\bibfnamefont {Marcel}\ \bibnamefont {Franz}}, \ and\
  \bibinfo {author} {\bibfnamefont {Ian}\ \bibnamefont {Affleck}},\ }\bibfield
  {title} {\enquote {\bibinfo {title} {Phase diagram of the interacting
  majorana chain model},}\ }\href@noop {} {\bibfield  {journal} {\bibinfo
  {journal} {Physical Review B}\ }\textbf {\bibinfo {volume} {92}},\ \bibinfo
  {pages} {235123} (\bibinfo {year} {2015})}\BibitemShut {NoStop}%
\bibitem [{\citenamefont {Leviatan}\ \emph {et~al.}()\citenamefont {Leviatan},
  \citenamefont {Pollmann}, \citenamefont {Bardarson}, \citenamefont {Huse},\
  and\ \citenamefont {Altman}}]{leviatanQuantumThermalizationDynamics2017}%
  \BibitemOpen
  \bibfield  {author} {\bibinfo {author} {\bibfnamefont {Eyal}\ \bibnamefont
  {Leviatan}}, \bibinfo {author} {\bibfnamefont {Frank}\ \bibnamefont
  {Pollmann}}, \bibinfo {author} {\bibfnamefont {Jens~H.}\ \bibnamefont
  {Bardarson}}, \bibinfo {author} {\bibfnamefont {David~A.}\ \bibnamefont
  {Huse}}, \ and\ \bibinfo {author} {\bibfnamefont {Ehud}\ \bibnamefont
  {Altman}},\ }\href {\doibase 10.48550/arXiv.1702.08894} {\enquote {\bibinfo
  {title} {Quantum thermalization dynamics with {{Matrix-Product States}}},}\
  }\Eprint {http://arxiv.org/abs/1702.08894} {1702.08894} \BibitemShut
  {NoStop}%
\bibitem [{\citenamefont {Parker}\ \emph {et~al.}(2019)\citenamefont {Parker},
  \citenamefont {Cao}, \citenamefont {Avdoshkin}, \citenamefont {Scaffidi},\
  and\ \citenamefont {Altman}}]{parkerUniversalOperatorGrowth2019}%
  \BibitemOpen
  \bibfield  {author} {\bibinfo {author} {\bibfnamefont {Daniel~E.}\
  \bibnamefont {Parker}}, \bibinfo {author} {\bibfnamefont {Xiangyu}\
  \bibnamefont {Cao}}, \bibinfo {author} {\bibfnamefont {Alexander}\
  \bibnamefont {Avdoshkin}}, \bibinfo {author} {\bibfnamefont {Thomas}\
  \bibnamefont {Scaffidi}}, \ and\ \bibinfo {author} {\bibfnamefont {Ehud}\
  \bibnamefont {Altman}},\ }\bibfield  {title} {\enquote {\bibinfo {title} {A
  {Universal} {Operator} {Growth} {Hypothesis}},}\ }\href {\doibase
  10.1103/PhysRevX.9.041017} {\bibfield  {journal} {\bibinfo  {journal}
  {Physical Review X}\ }\textbf {\bibinfo {volume} {9}},\ \bibinfo {pages}
  {041017} (\bibinfo {year} {2019})},\ \bibinfo {note} {arXiv:
  1812.08657}\BibitemShut {NoStop}%
\bibitem [{\citenamefont {Mukerjee}\ \emph {et~al.}(2006)\citenamefont
  {Mukerjee}, \citenamefont {Oganesyan},\ and\ \citenamefont
  {Huse}}]{mukerjeeStatisticalTheoryTransport2006}%
  \BibitemOpen
  \bibfield  {author} {\bibinfo {author} {\bibfnamefont {Subroto}\ \bibnamefont
  {Mukerjee}}, \bibinfo {author} {\bibfnamefont {Vadim}\ \bibnamefont
  {Oganesyan}}, \ and\ \bibinfo {author} {\bibfnamefont {David}\ \bibnamefont
  {Huse}},\ }\bibfield  {title} {\enquote {\bibinfo {title} {Towards a
  statistical theory of transport by strongly-interacting lattice fermions},}\
  }\href {\doibase 10.1103/PhysRevB.73.035113} {\bibfield  {journal} {\bibinfo
  {journal} {Physical Review B}\ }\textbf {\bibinfo {volume} {73}} (\bibinfo
  {year} {2006}),\ 10.1103/PhysRevB.73.035113},\ \Eprint
  {http://arxiv.org/abs/cond-mat/0503177} {arxiv:cond-mat/0503177} \BibitemShut
  {NoStop}%
\bibitem [{\citenamefont
  {Delacretaz}(2020)}]{delacretazHeavyOperatorsHydrodynamic2020}%
  \BibitemOpen
  \bibfield  {author} {\bibinfo {author} {\bibfnamefont {Luca~V.}\ \bibnamefont
  {Delacretaz}},\ }\bibfield  {title} {\enquote {\bibinfo {title} {Heavy
  {{Operators}} and {{Hydrodynamic Tails}}},}\ }\href {\doibase
  10.21468/SciPostPhys.9.3.034} {\bibfield  {journal} {\bibinfo  {journal}
  {SciPost Physics}\ }\textbf {\bibinfo {volume} {9}},\ \bibinfo {pages} {034}
  (\bibinfo {year} {2020})},\ \Eprint {http://arxiv.org/abs/2006.01139}
  {arxiv:2006.01139} \BibitemShut {NoStop}%
\bibitem [{\citenamefont {Michailidis}\ \emph {et~al.}(2023)\citenamefont
  {Michailidis}, \citenamefont {Abanin},\ and\ \citenamefont
  {Delacrétaz}}]{michailidisCorrectionsDiffusionInteracting2023}%
  \BibitemOpen
  \bibfield  {author} {\bibinfo {author} {\bibfnamefont {Alexios~A.}\
  \bibnamefont {Michailidis}}, \bibinfo {author} {\bibfnamefont {Dmitry~A.}\
  \bibnamefont {Abanin}}, \ and\ \bibinfo {author} {\bibfnamefont {Luca~V.}\
  \bibnamefont {Delacrétaz}},\ }\href {http://arxiv.org/abs/2310.10564}
  {\enquote {\bibinfo {title} {Corrections to diffusion in interacting quantum
  systems},}\ } (\bibinfo {year} {2023}),\ \bibinfo {note} {arXiv:2310.10564
  [cond-mat, physics:hep-th]}\BibitemShut {NoStop}%
\bibitem [{\citenamefont {Yoo}\ \emph {et~al.}(2023)\citenamefont {Yoo},
  \citenamefont {White},\ and\ \citenamefont
  {Swingle}}]{yooOpensystemSpinTransport2023}%
  \BibitemOpen
  \bibfield  {author} {\bibinfo {author} {\bibfnamefont {Yongchan}\
  \bibnamefont {Yoo}}, \bibinfo {author} {\bibfnamefont {Christopher~David}\
  \bibnamefont {White}}, \ and\ \bibinfo {author} {\bibfnamefont {Brian}\
  \bibnamefont {Swingle}},\ }\bibfield  {title} {\enquote {\bibinfo {title}
  {Open-system spin transport and operator weight dissipation in spin
  chains},}\ }\href {\doibase 10.1103/PhysRevB.107.115118} {\bibfield
  {journal} {\bibinfo  {journal} {Physical Review B}\ }\textbf {\bibinfo
  {volume} {107}},\ \bibinfo {pages} {115118} (\bibinfo {year}
  {2023})}\BibitemShut {NoStop}%
\bibitem [{\citenamefont {Surace}\ and\ \citenamefont
  {Motrunich}(2023{\natexlab{b}})}]{surace2023weak}%
  \BibitemOpen
  \bibfield  {author} {\bibinfo {author} {\bibfnamefont {Federica~Maria}\
  \bibnamefont {Surace}}\ and\ \bibinfo {author} {\bibfnamefont {Olexei}\
  \bibnamefont {Motrunich}},\ }\bibfield  {title} {\enquote {\bibinfo {title}
  {Weak integrability breaking perturbations of integrable models},}\
  }\href@noop {} {\bibfield  {journal} {\bibinfo  {journal} {arXiv preprint
  arXiv:2302.12804}\ } (\bibinfo {year} {2023}{\natexlab{b}})}\BibitemShut
  {NoStop}%
\bibitem [{\citenamefont {Stoudenmire}\ and\ \citenamefont
  {White}(2010)}]{Stoudenmire_2010}%
  \BibitemOpen
  \bibfield  {author} {\bibinfo {author} {\bibfnamefont {E~M}\ \bibnamefont
  {Stoudenmire}}\ and\ \bibinfo {author} {\bibfnamefont {Steven~R}\
  \bibnamefont {White}},\ }\bibfield  {title} {\enquote {\bibinfo {title}
  {Minimally entangled typical thermal state algorithms},}\ }\href {\doibase
  10.1088/1367-2630/12/5/055026} {\bibfield  {journal} {\bibinfo  {journal}
  {New Journal of Physics}\ }\textbf {\bibinfo {volume} {12}},\ \bibinfo
  {pages} {055026} (\bibinfo {year} {2010})}\BibitemShut {NoStop}%
\bibitem [{\citenamefont {Paeckel}\ \emph {et~al.}(2019)\citenamefont
  {Paeckel}, \citenamefont {Köhler}, \citenamefont {Swoboda}, \citenamefont
  {Manmana}, \citenamefont {Schollwöck},\ and\ \citenamefont
  {Hubig}}]{PAECKEL2019167998}%
  \BibitemOpen
  \bibfield  {author} {\bibinfo {author} {\bibfnamefont {Sebastian}\
  \bibnamefont {Paeckel}}, \bibinfo {author} {\bibfnamefont {Thomas}\
  \bibnamefont {Köhler}}, \bibinfo {author} {\bibfnamefont {Andreas}\
  \bibnamefont {Swoboda}}, \bibinfo {author} {\bibfnamefont {Salvatore~R.}\
  \bibnamefont {Manmana}}, \bibinfo {author} {\bibfnamefont {Ulrich}\
  \bibnamefont {Schollwöck}}, \ and\ \bibinfo {author} {\bibfnamefont
  {Claudius}\ \bibnamefont {Hubig}},\ }\bibfield  {title} {\enquote {\bibinfo
  {title} {Time-evolution methods for matrix-product states},}\ }\href
  {\doibase https://doi.org/10.1016/j.aop.2019.167998} {\bibfield  {journal}
  {\bibinfo  {journal} {Annals of Physics}\ }\textbf {\bibinfo {volume}
  {411}},\ \bibinfo {pages} {167998} (\bibinfo {year} {2019})}\BibitemShut
  {NoStop}%
\bibitem [{\citenamefont {Barthel}\ and\ \citenamefont
  {Zhang}(2020)}]{barthel2020optimized}%
  \BibitemOpen
  \bibfield  {author} {\bibinfo {author} {\bibfnamefont {Thomas}\ \bibnamefont
  {Barthel}}\ and\ \bibinfo {author} {\bibfnamefont {Yikang}\ \bibnamefont
  {Zhang}},\ }\bibfield  {title} {\enquote {\bibinfo {title} {Optimized
  lie--trotter--suzuki decompositions for two and three non-commuting terms},}\
  }\href@noop {} {\bibfield  {journal} {\bibinfo  {journal} {Annals of
  Physics}\ }\textbf {\bibinfo {volume} {418}},\ \bibinfo {pages} {168165}
  (\bibinfo {year} {2020})}\BibitemShut {NoStop}%
\bibitem [{Note1()}]{Note1}%
  \BibitemOpen
  \bibinfo {note} {ITensor svd keywords use\protect \_relative\protect
  \_cutoff=true, cutoff=1e-8}\BibitemShut {NoStop}%
\bibitem [{\citenamefont {Noh}\ \emph {et~al.}(2020)\citenamefont {Noh},
  \citenamefont {Jiang},\ and\ \citenamefont
  {Fefferman}}]{Noh2020efficientclassical}%
  \BibitemOpen
  \bibfield  {author} {\bibinfo {author} {\bibfnamefont {Kyungjoo}\
  \bibnamefont {Noh}}, \bibinfo {author} {\bibfnamefont {Liang}\ \bibnamefont
  {Jiang}}, \ and\ \bibinfo {author} {\bibfnamefont {Bill}\ \bibnamefont
  {Fefferman}},\ }\bibfield  {title} {\enquote {\bibinfo {title} {Efficient
  classical simulation of noisy random quantum circuits in one dimension},}\
  }\href {\doibase 10.22331/q-2020-09-11-318} {\bibfield  {journal} {\bibinfo
  {journal} {{Quantum}}\ }\textbf {\bibinfo {volume} {4}},\ \bibinfo {pages}
  {318} (\bibinfo {year} {2020})}\BibitemShut {NoStop}%
\bibitem [{\citenamefont {Li}\ \emph {et~al.}(2019)\citenamefont {Li},
  \citenamefont {Wang},\ and\ \citenamefont {Cai}}]{li2019tutorial}%
  \BibitemOpen
  \bibfield  {author} {\bibinfo {author} {\bibfnamefont {Xiaocan}\ \bibnamefont
  {Li}}, \bibinfo {author} {\bibfnamefont {Shuo}\ \bibnamefont {Wang}}, \ and\
  \bibinfo {author} {\bibfnamefont {Yinghao}\ \bibnamefont {Cai}},\ }\bibfield
  {title} {\enquote {\bibinfo {title} {Tutorial: Complexity analysis of
  singular value decomposition and its variants},}\ }\href@noop {} {\bibfield
  {journal} {\bibinfo  {journal} {arXiv preprint arXiv:1906.12085}\ } (\bibinfo
  {year} {2019})}\BibitemShut {NoStop}%
\bibitem [{\citenamefont {Lloyd}\ \emph {et~al.}(2023)\citenamefont {Lloyd},
  \citenamefont {Rakovszky}, \citenamefont {Pollmann},\ and\ \citenamefont {von
  Keyserlingk}}]{lloydBallisticDiffusiveCrossover2023}%
  \BibitemOpen
  \bibfield  {author} {\bibinfo {author} {\bibfnamefont {Jerome}\ \bibnamefont
  {Lloyd}}, \bibinfo {author} {\bibfnamefont {Tibor}\ \bibnamefont
  {Rakovszky}}, \bibinfo {author} {\bibfnamefont {Frank}\ \bibnamefont
  {Pollmann}}, \ and\ \bibinfo {author} {\bibfnamefont {Curt}\ \bibnamefont
  {von Keyserlingk}},\ }\href {http://arxiv.org/abs/2310.16043} {\enquote
  {\bibinfo {title} {The ballistic to diffusive crossover in a
  weakly-interacting {Fermi} gas},}\ } (\bibinfo {year} {2023}),\ \bibinfo
  {note} {arXiv:2310.16043 [cond-mat, physics:quant-ph]}\BibitemShut {NoStop}%
\bibitem [{\citenamefont
  {Vidal}(2003)}]{vidalEfficientClassicalSimulation2003}%
  \BibitemOpen
  \bibfield  {author} {\bibinfo {author} {\bibfnamefont {Guifre}\ \bibnamefont
  {Vidal}},\ }\bibfield  {title} {\enquote {\bibinfo {title} {Efficient
  classical simulation of slightly entangled quantum computations},}\ }\href
  {\doibase 10.1103/PhysRevLett.91.147902} {\bibfield  {journal} {\bibinfo
  {journal} {Physical Review Letters}\ }\textbf {\bibinfo {volume} {91}}
  (\bibinfo {year} {2003}),\ 10.1103/PhysRevLett.91.147902},\ \Eprint
  {http://arxiv.org/abs/quant-ph/0301063} {arxiv:quant-ph/0301063} \BibitemShut
  {NoStop}%
\bibitem [{\citenamefont
  {Vidal}(2004)}]{vidalEfficientSimulationOnedimensional2004}%
  \BibitemOpen
  \bibfield  {author} {\bibinfo {author} {\bibfnamefont {G.}~\bibnamefont
  {Vidal}},\ }\bibfield  {title} {\enquote {\bibinfo {title} {Efficient
  simulation of one-dimensional quantum many-body systems},}\ }\href {\doibase
  10.1103/PhysRevLett.93.040502} {\bibfield  {journal} {\bibinfo  {journal}
  {Physical Review Letters}\ }\textbf {\bibinfo {volume} {93}} (\bibinfo {year}
  {2004}),\ 10.1103/PhysRevLett.93.040502},\ \Eprint
  {http://arxiv.org/abs/quant-ph/0310089} {arxiv:quant-ph/0310089} \BibitemShut
  {NoStop}%
\bibitem [{\citenamefont {Zwolak}\ and\ \citenamefont
  {Vidal}(2004)}]{zwolakMixedstateDynamicsOnedimensional2004}%
  \BibitemOpen
  \bibfield  {author} {\bibinfo {author} {\bibfnamefont {Michael}\ \bibnamefont
  {Zwolak}}\ and\ \bibinfo {author} {\bibfnamefont {Guifre}\ \bibnamefont
  {Vidal}},\ }\bibfield  {title} {\enquote {\bibinfo {title} {Mixed-state
  dynamics in one-dimensional quantum lattice systems: A time-dependent
  superoperator renormalization algorithm},}\ }\href {\doibase
  10.1103/PhysRevLett.93.207205} {\bibfield  {journal} {\bibinfo  {journal}
  {Physical Review Letters}\ }\textbf {\bibinfo {volume} {93}} (\bibinfo {year}
  {2004}),\ 10.1103/PhysRevLett.93.207205},\ \Eprint
  {http://arxiv.org/abs/cond-mat/0406440} {arxiv:cond-mat/0406440} \BibitemShut
  {NoStop}%
\end{thebibliography}%

\appendix

\section{Matrix product operator representation of fermionic DAOE}\label{app:FDAOE-mpo}

In this appendix, we give more details on the construction of the MPO representation of the fermionic DAOE superoperator $\mathcal{M}_{w_{*},\gamma}$.
As explained in \cref{s:superoperator}, this construction relies on the Jordan-Wigner embedding of fermionic operators into the space of operators of qubits. This mapping
maps each Pauli string -- a single product of Pauli operators or the identity operator on each site -- to a corresponding product of Majorana operators with 0, 1, or 2 Majorana factors at each site, up to a phase factor.
Similarly, every product of Majorana operators maps to a Pauli string. As the MPO representation of the superoperator is by construction a linear superoperator, and because the Pauli string operators form a basis for the full space of operators, it is sufficient to confirm that the MPO representation has the correct action on each Pauli string.

The MPO is constructed with a constant rank-4 tensor $W^{nn'}_{ab}$, where $n, n' \in \{ I, X, Y, Z \}.$ To use a convenient notation, we will allow the virtual indices to take values of the form $a_s \in \{ 0_+, 0_-, 1_+, 1_-, \ldots w^*_+, w^*_- \},$ where the integer part of the label $a$ is used to track a fermion weight and the subscript $s$ is used to track a fermion parity. There are $2 (w^*+1)$ such labels --- however, we will find that we only use the labels $0_+, 1_-, 2_+, 3_-, \ldots$ and $w^*_+, w^*_-$, where for $a<w^*$ the parity label $s$ matches the parity of $a$ as an integer. This results in a set of $w^*+2$ total labels, and thus the bond dimension of our MPO representation is $w^*+2$.

The non-zero matrix elements of our MPO tensor $W^{nn'}_{ab}$ are as follows:
\begin{align}
    W^{II}_{a_+,b_+}&=W^{ZZ}_{a_-,b_-}=\delta_{a,b} \\
    W^{XX}_{a_+,b_-}&=W^{YY}_{a_+,b_-}=\delta_{a+1,b}+e^{-\gamma}\delta_{a,w^*}\delta_{b,w^*} \nonumber \\
    W^{XX}_{a_-,b_+}&=W^{YY}_{a_-,b_+}=\delta_{a+1,b}+e^{-\gamma}\delta_{a,w^*}\delta_{b,w^*} \nonumber \\
    W^{II}_{a_-,b_-}&=W^{ZZ}_{a_+,b_+} =\delta_{a+2,b} + e^{-2\gamma}\delta_{a, w^*}\delta_{b, w^*}\nonumber \\
    &\qquad\qquad\qquad + e^{-\gamma}\delta_{a, w^*-1}\delta_{b, w^*} \nonumber
    \label{eq:MPOtensor}
\end{align}
The left and right most tensor in the MPO representation are to be contracted with vectors $v^L_a = \delta_{a, 0_+}$ and $v^R_b = 1$ on the left and right virtual bond, respectively. 

To understand this implementation of the MPO, we first note that the non-zero elements listed guarantee a consistent tracking of fermion parity. In more exact terms,
the only contributions to $M_{w^*, \gamma}[O]$ for a Pauli string $O = \prod_i O_i$ occur when the parity index of the bond between site $j$ and $j+1$ matches
the fermion parity of $\prod_{i \leq j} O_i$ for all $j$. This is convenient, as matching the fermion parity allow us to determine whether the basis operator $O$ has an even or odd number of Jordan-Wigner string factors from \cref{eq:sigma-as-majorana} that cross the bond $j \to j+1$.

\begin{figure}[t]
\includegraphics[width=0.5\textwidth]{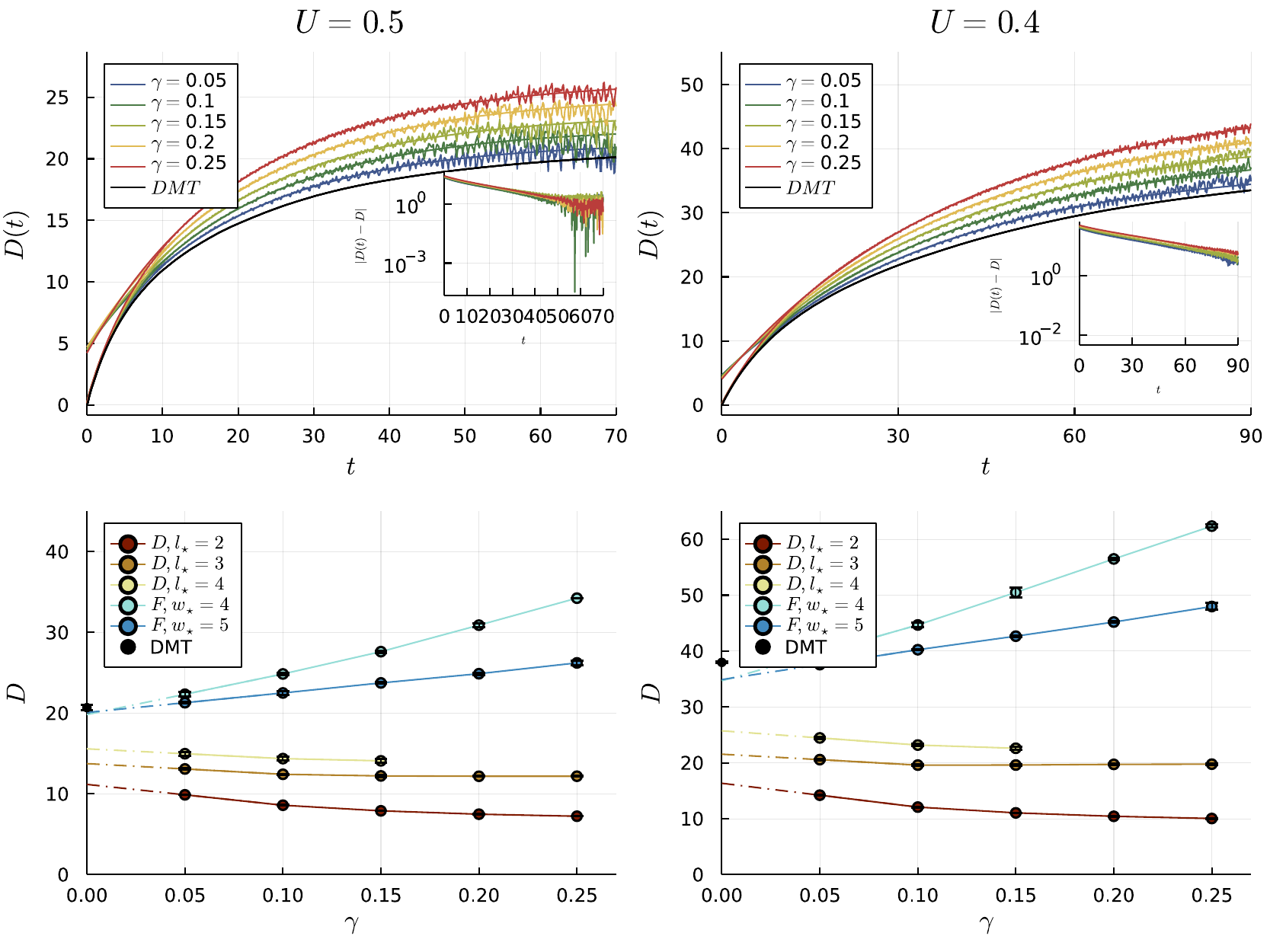}
\caption{Analogues of \ref{fig:Dt-03} and \ref{fig:Dinf-gamma} for $U=0.4$ and $U=0.5$.
The left panel is $U=0.5$ and the right one is $U=0.4$. All data comes from Trotter step $dt =0.1$, and cutoff $\epsilon=10^{-8}$.}
\label{fig:D-gamma-other-U}
\end{figure}

Similarly, we can see that at all bonds the fermion weight of the virtual index must match either the fermion weight of $\prod_{i \leq j} O_i$ or $w^*$, whichever is smaller. When there are an even number of Jordan-Wigner string factors, the presence of an $I$ in a Pauli string corresponds to a Majorana product with 0 Majorana operators on that site; if instead there are an odd number, the presence of $Z$ does instead. This gives the first line of \cref{eq:MPOtensor}. The second line corresponds to fermion weight $1$ operators, which either increases the fermion weight counter by $1$ or keeps it constant if it has reached the maximum $w^*$. The $e^{-\gamma}$ factor dissipates Pauli strings for each additional unit of fermion weight beyond $w^*$. Finally, in the last line of \cref{eq:MPOtensor}, we have Pauli string factors that correspond to fermion weight $2$ operators, which requires increasing the fermion weight counter twice or increasing it to $w^*$ and dissipating the operator by a factor of $e^{-\gamma}$ or $e^{-2\gamma}$, depending on whether the fermion weight goes 1 or 2 units beyond $w^*$.

\section{$D(t)$ and extrapolation for other interaction strengths}\label{app:other-interaction}

Fig.~\ref{fig:D-gamma-other-U} shows analogues of Fig's \ref{fig:Dt-03} and \ref{fig:Dinf-gamma} for $U = 0.4$ and $U = 0.5$.
Fig.~\ref{fig:D-gamma-other-U} top shows $D(t)$, together with fits and (in the inset) a logarithmic difference from the $D = \lim_{t \to \infty} D(t)$ resulting from the fit, across $\gamma$.
Fig.~\ref{fig:D-gamma-other-U} bottom shows $D$ from fit as a function of $\gamma$, together with the extrapolation from the last two points.
Table

\begin{table}[h!]
\centering
\begin{tabular}{|lllll|}
\hline
$U$ & $\delta t$ & $\chi_{\max}$ & $D$ (mean) & $D$ (std) \\
\hline
0.3 & 0.0625 & 128 & 96.8 & 5.7 \\
0.3 & 0.125 & 128 & 106.238 & 14.6 \\
0.3 & 0.125 & 256 & 102.367 & 2.38278 \\
\hline
0.4 &  0.0625 & 128  &  39.4765  &  2.01023\\
0.4 &  0.0625 & 256  &  38.2516  &  0.703325\\
0.4 &  0.125  & 128  &  36.3361  &  0.853394\\
0.4 &  0.125  & 256  &  37.9766  &  0.126748\\
\hline
0.5 &  0.0625 & 128  &  21.1115  &  1.213\\
0.5 &  0.0625 & 256  &  21.3559  &  0.333348\\
0.5 &  0.125  & 128  &  20.1833  &  0.896041\\
0.5 &  0.125  & 256  &  20.6832  &  0.359822\\
\hline
\end{tabular}
\caption{Diffusion coefficients from fit to DMT $D(t)$. Mean and standard deviation are across three different fit-window end times.}
\label{table:DMT}
\end{table}

\begin{table}[h!]
\centering
\begin{tabular}{|llll|}
\hline
$U$ & $w_*$ & $D$ (mean) & $D$ (std) \\
0.3  & 4.0  &  94.9121 & 4.37422 \\
0.3  & 5.0  &  89.4662 & 2.67036 \\
\hline
0.35 & 4.0  &  59.2131 & 3.09694 \\
0.35 & 5.0  &  51.6491 & 1.63505 \\
\hline
0.4  & 4.0  &  36.2878 & 1.60103 \\
0.4  & 5.0  &  35.2506 & 4.91012 \\
\hline
0.45 & 4.0  &  24.3317 & 1.3542 \\
0.45 & 5.0  &  25.2716 & 1.03157 \\
\hline
0.5  & 4.0  &  19.8595 & 2.73888 \\
0.5  & 5.0  &  20.9656 & 0.495984 \\
\hline
0.55 & 4.0  &  16.4897 & 0.502595 \\
0.55 & 5.0  &  15.1526 & 0.807459 \\
\hline
0.6  & 4.0  &  14.0067 & 0.270231 \\
0.6  & 5.0  &  13.8689 & 0.562507 \\
\hline
\end{tabular}

\caption{Diffusion coefficients from fit to FDAOE $D(t)$. Mean and standard deviation are across three different fit-window end times.}
\label{table:FDAOE}
\end{table}

\section{Convergence in SVD cutoff $\epsilon$}\label{app:epsilon-convergence}

\begin{figure}[t]
    \centering
    \includegraphics[scale=0.7]{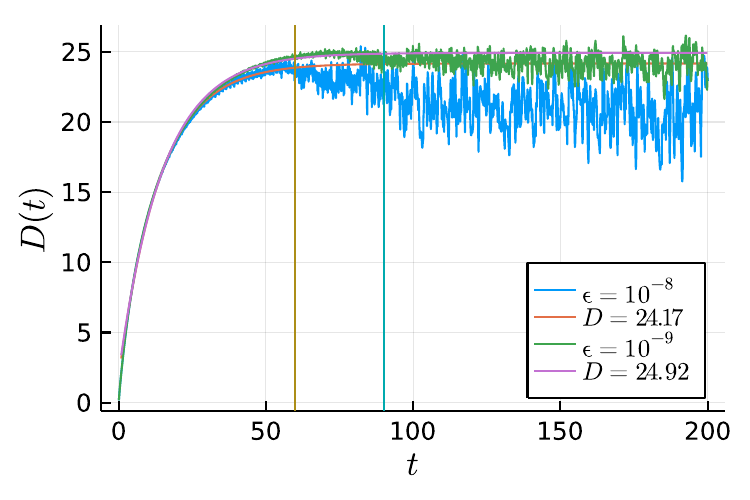}
    \caption{We compared $D(t)$ with two truncation errors $10^{-8}$ and $10^{-9}$ with $U=0.55, w_{*}=4,\gamma=0.2$. We also do the exponential extrapolation and see that $D$ only differs less than $5$ percent.}
    \label{fig:tr}
\end{figure}

\begin{figure}
    \includegraphics[width=0.95\columnwidth]{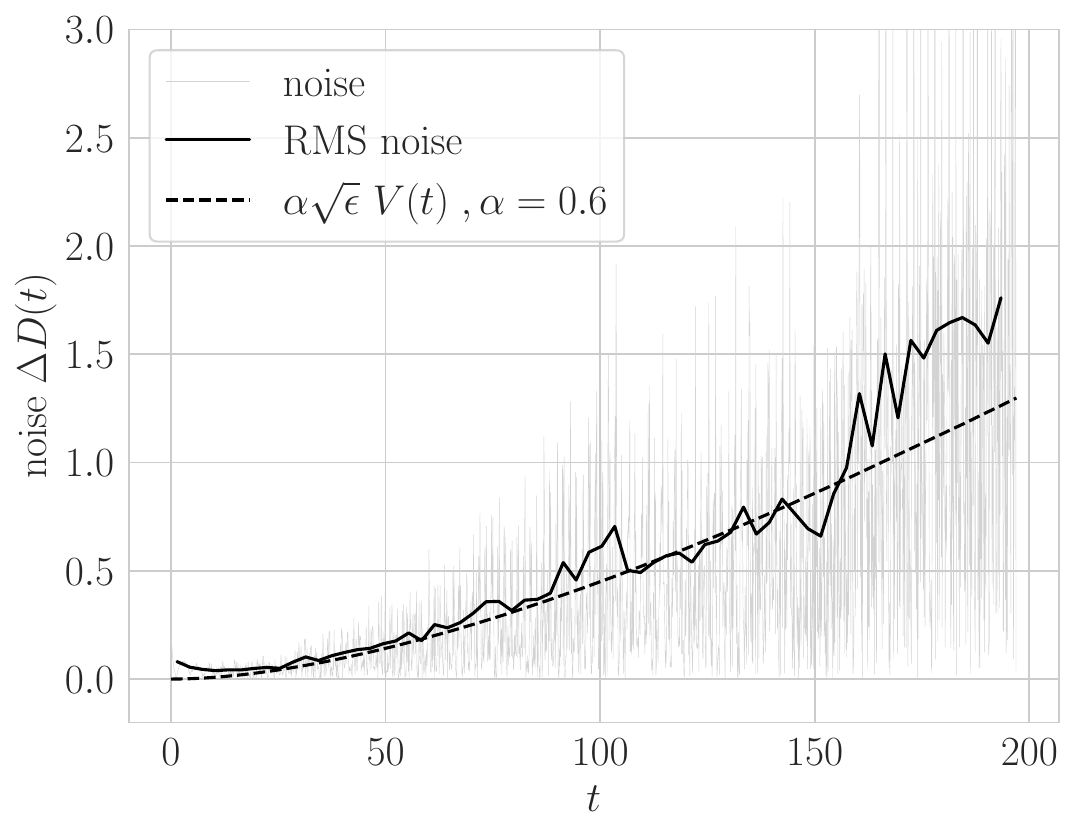}
    \includegraphics[width=0.95\columnwidth]{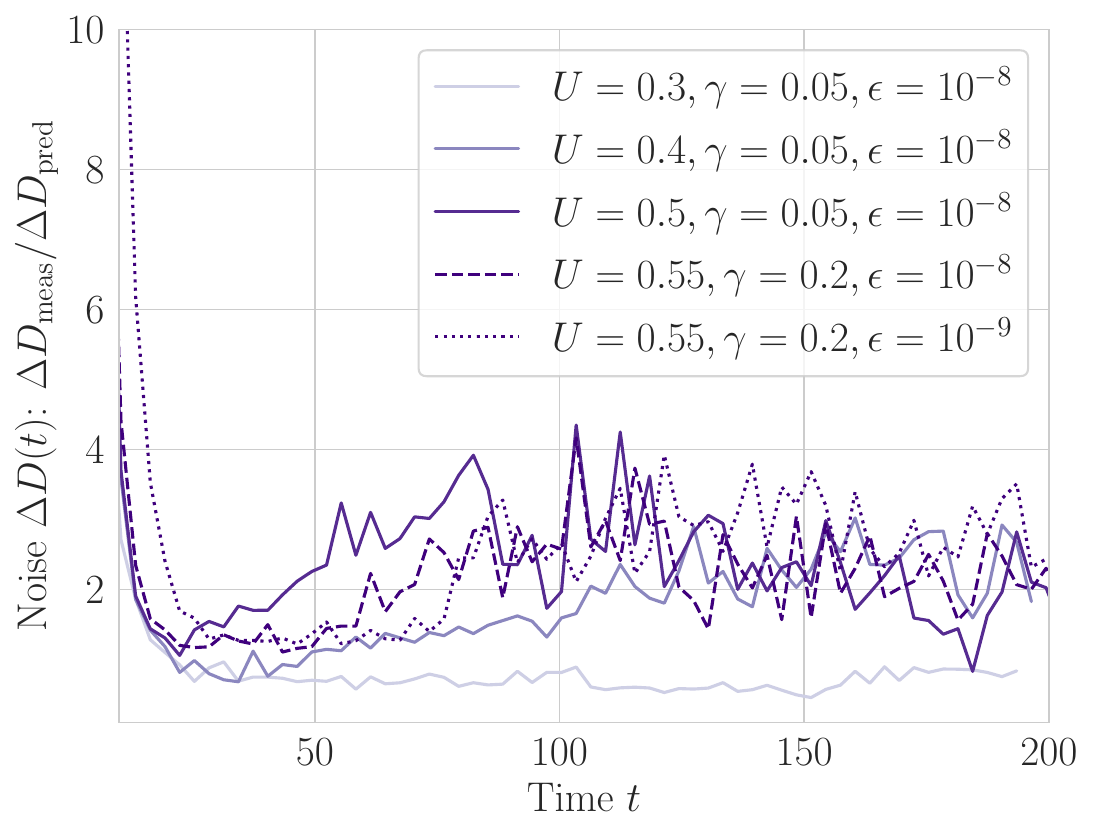}
    \caption{\textbf{Top}: Noise magnitude at $U =0.3$, $\gamma = 0.05$, $\epsilon = 10^{-8}$ compared to the prediction $\Delta D_{\mathrm{pred}}(t) = \alpha \sqrt{\epsilon}\, V(t)$\ \eqref{eq:Dt-noise-prediction}, with the fit parameter $\alpha$ chosen by eye.
    \textbf{Bottom}: noise at a variety of $U, \gamma$, and $\epsilon$ normalized by the prediction $\Delta D_{\mathrm{pred}}$ with $\alpha = 1$.
    }
    \label{fig:noise-mag-pred}
\end{figure}

In simulating dynamics with FDAOE we apply Trotter gates and the FDAOE MPO;
after each application we discard small singular values
\begin{align}
\sum_{\alpha\text{ discarded}} s_\alpha^2 < \epsilon \left[ \sum_\beta s_\beta^2\right] \;.
\end{align}
Fig.~\ref{fig:tr} shows $D(t)$ for $\epsilon = 10^{-8}, 10^{-9}$ at $U = 0.55, \gamma = 0.2$.
The noise in each $D(t)$ worsens with time, and it is smaller for $\epsilon = 10^{-9}$ than for $\epsilon = 10^{-8}$.
Fig.~\ref{fig:noise-mag-pred} top shows the RMS noise as a function of time for $U = 0.3$, $\gamma = 0.05$, $\epsilon = 10^{-8}$;
there we see that it is in fact roughly proportional to the mean square displacement $V(t)$.
(We describe how we calculate the noise in App.~\ref{app:ss:noise-meas} below.)
A heuristic a priori argument (App.~\ref{app:ss:noise-pred} below) predicts a noise magnitude
\begin{align}
\label{eq:Dt-noise-prediction}
    \Delta D_{\mathrm{pred}}(t) = \alpha \sqrt{ \epsilon }\ V(t)\;,
\end{align}
where $\alpha$ is a fit parameter depending (in part) on the timestep $\delta t$.

Fig.~\ref{fig:noise-mag-pred} shows the ratio of the measured noise magnitude $\Delta D_{\mathrm{meas}}$ to the prediction $\Delta D_{\mathrm{pred}}$ of \eqref{eq:Dt-noise-prediction} for a variety of $U, \gamma$, and $\epsilon$.
For short times ($t \lesssim 25$) the ratio is large.
This is in part because the prediction $\Delta D_{\mathrm meas}$ is initially small,
because the mean square displacement $V(t)$ is small.
Additionally the measured noise displays a small peak at $t = 0$, already visible in Fig.~\ref{fig:noise-mag-pred} top,
resulting from the details of our noise measurement procedure.
For $t \gtrsim 25$, the noise magnitude is reasonably well-predicted by \eqref{eq:Dt-noise-prediction}.

\subsection{Heuristic a priori estimate of noise in $D(t)$ due to SVD truncation}\label{app:ss:noise-pred}
Heuristically, the truncation applies a random perturbation of magnitude $\sqrt{\epsilon}$ to the operator truncated.
When the operator truncated is the Heisenberg operator $\varepsilon_{L/2}(t)$,
truncation maps
\begin{align}
    \varepsilon_{L/2}(t) \mapsto (1 + \sqrt \epsilon W)[\varepsilon_{L/2}(t)]\;,
\end{align}
where $V$ is some (not necessarily unitary) superoperator;
this changes the correlation function to
\begin{align}
\begin{split}
    C^{\varepsilon\varepsilon}(x,t) &= \tr\left(\varepsilon_x(1 + \sqrt \epsilon W)[\varepsilon_{L/2}(t)]\right) \\
    &= C^{\varepsilon\varepsilon}(x,t) + \sqrt\epsilon\; \tr\Big(\varepsilon_x W [\varepsilon_{L/2}(t)]\Big)\;.
\end{split}
\end{align}
The superoperator $W$ acts locally.
To understand this, recall that $\varepsilon_{L/2}(t)$ is a low-bond dimension MPO,
so it has a correlation length set by the leading nontrivial eigenvalue of the transfer matrix.
Perturbations like truncation heal within that correlation length,
so the superoperator $W$ acts with a range given by that correlation length.

Since $W$ acts locally, estimate
\begin{align}
\begin{split}
    \tr[\varepsilon_x W \varepsilon_{L/2}(t)] &= \xi(x,t) \tr[\varepsilon_x \varepsilon_{L/2}(t)] \\
    &\equiv \xi(x,t) \Cee(x,t)
\end{split}
\end{align}
where $\xi(x,t)$ is a random variable with 
\begin{align}
    \expct{\xi(x,t) \xi(x',t')} = \alpha^2 \delta_{xx'}\delta(t-t')\;,
\end{align}
$\alpha$ some constant.
Truncation then takes
\begin{align}
    \Cee \mapsto [1 + \sqrt{\epsilon}\; \xi(x,t)]\Cee(x,t)\;.
\end{align}
and the mean squared displacement
\begin{align}
\begin{split}
    V(t) &\mapsto V'(t) = \sum_x x^2 [1 + \sqrt{\epsilon}\; \xi(x,t)]\Cee(x,t)\\
   &= V(t) + \Delta V_{trunc}(t)\;,
\end{split}
\end{align}
with
\begin{align}
    \Delta V_{trunc}(t) = \sqrt \epsilon \sum_x \; x^2 \Cee(x,t) \xi(x,t)\;.
\end{align}
This truncation appears as noise in the time-dependent diffusion coefficient:
the numerical derivative leading to the diffusion coefficient is
\begin{align}
   D(t) &= \frac 1 2 \delta t^{-1} [ V'(t+\delta t) - V(\delta t)] \\
   &= D_{phys} +  \frac 1 2 \delta t^{-1} \Delta V_{trunc}(t)
\end{align}
where $D_{phys}$ is the ``physical'' contribution to the numerical derivative, coming from the pre-truncation timestep,
and the second term $\propto \Delta V_{trunc}(t)$ is the noise coming from the truncation.
We can then estimate the magnitude of the noise by treating $\xi(x,t)$, hence $\Delta V_{trunc}$, as random variables
and estimating the variance:
\begin{align}
\begin{split}
   \expct{\Delta V_{trunc}(t)^2} &= \epsilon \expct{ \left[\sum_x \; x^2 \Cee(x,t) \xi(x,t)\right]^2 }  \\
   &=\epsilon \sum_x x^4 \Cee(x,t) \\
   &= \alpha \epsilon V(t)^2
\end{split}
\end{align}
using  $\expct{\xi(x,t) \xi(x',t')} = \alpha \delta_{xx'}\delta(t-t')$
and sweeping some dimensionless factors into $\alpha$.
The standard deviation of the noise in $D(t)$ is therefore
\begin{align}\label{eq:Dt-noise-prediction-2}
    \alpha \sqrt {\delta t^{-1} \epsilon/2} \ V(t)\;.
\end{align}
This expression includes a dependence on Trotter step $\delta t$ coming from the numerical derivative.
But the numerical derivative is not the only source of $\delta t$ dependence.
Consider, for example, the limit of small $\delta t$.
In that limit a Trotter step introduces only small Schmidt values,
which are all discarded by truncation:
that is, the truncation can undo the effect of time evolution.
We do not claim to consider all sources of $\delta t$-dependence, so we sweep it into the constant $\alpha$.
The predicted standard deviation of the noise is then
\begin{align}\label{eq:Dt-noise-prediction-2}
    \Delta D_{\mathrm{pred}} = \alpha \sqrt {\epsilon/2} \ V(t)\;.
\end{align}
Fig.~\ref{fig:noise-mag-pred} shows the noise compared to the prediction; we see reasonable agreement.

\subsection{Estimating the noise magnitude}\label{app:ss:noise-meas}

We seek to estimate the noise magnitude without reference to a global fit like the exponential fit of \eqref{s:results}, 
In brief, we estimate noise by binning in time, averaging $D(t)$ in each bin, constructing a linear interpolant between averages, mand measuring the RMS deviation from the interpolant.
In more detail, we
\begin{enumerate}
    \item compute variances $V_j$ at timesteps $(j-1)\delta t$, $j = 1 \dots$. (Throughout this section $\delta t = 0.1$.)
    \item\label{step:Djtj} compute time-dependent diffusion coefficients 
    \begin{align}
        D_j = \delta t^{-1} [V_{j+1} - V_j]\;;
    \end{align}
    assign them to times
    \begin{align}
        t_j = (j-1/2)\delta t\;.
    \end{align}
    \item Bin and average $D(t)$ over bins of width $n = 30$, corresponding to a time window $3$: that is, compute
    \begin{align}
    \begin{split}
        \bar D_k &= \frac 1 n \sum_{j = n(k-1)+1}^{nk} D_j\\
        &= (n\delta t)^{-1}[V_{nk+1} - V_{n(k-1)+1}]\;.
    \end{split}
    \end{align}
    Assign $\bar D_k$ to a time
    \begin{align}
        \bar t_k = (nk-1 -n/2)\delta t\;.
    \end{align}
    \item Form a linear interpolant $T(t)$ between the points $(\bar t_k, \bar D_k)$. (For $t < \bar t_1$ we linearly extrapolate.)
    \item Form errors 
    \begin{align}
        E_j = D_j - T(t_j)
    \end{align}
    with $D_j, t_j$ from step \ref{step:Djtj}.
    \item Take the RMS of $E_j$ over windows of 30 points, corresponding to time windows of size $3$, for $\Delta D_{\mathrm {meas}}$
\end{enumerate}

\section{DMT simulations}\label{app:DMT}
\begin{figure*}[t]
\includegraphics[width=\textwidth]{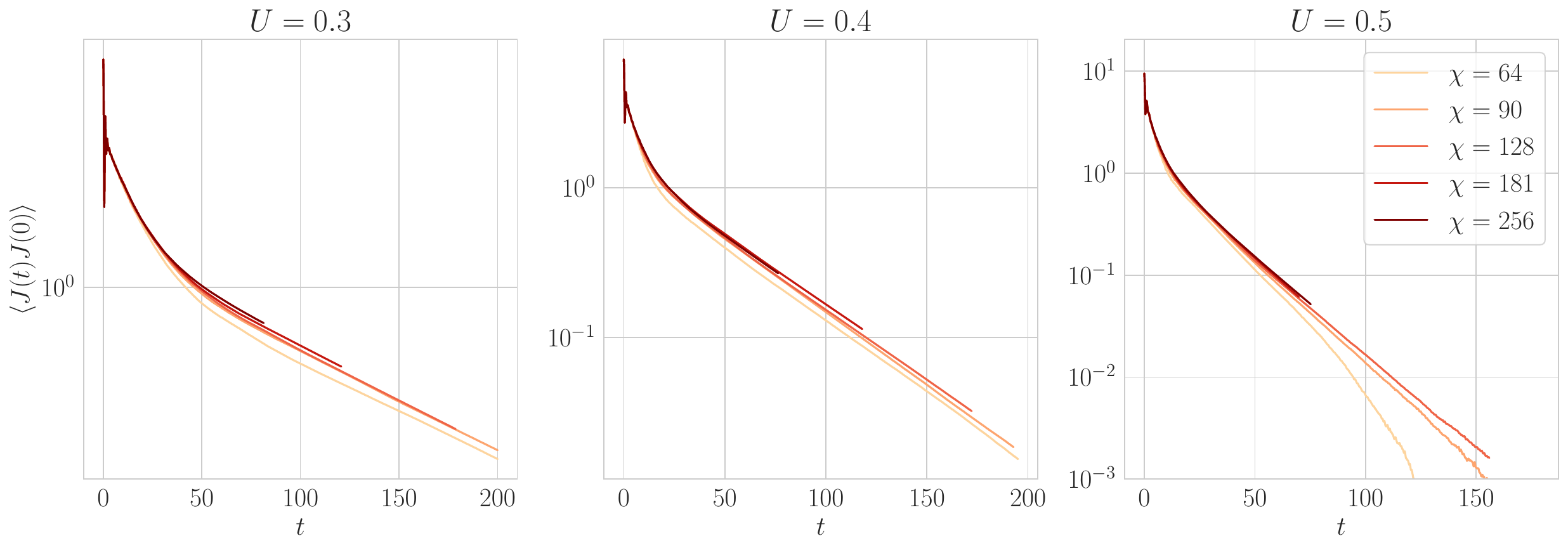}
\includegraphics[width=\textwidth]{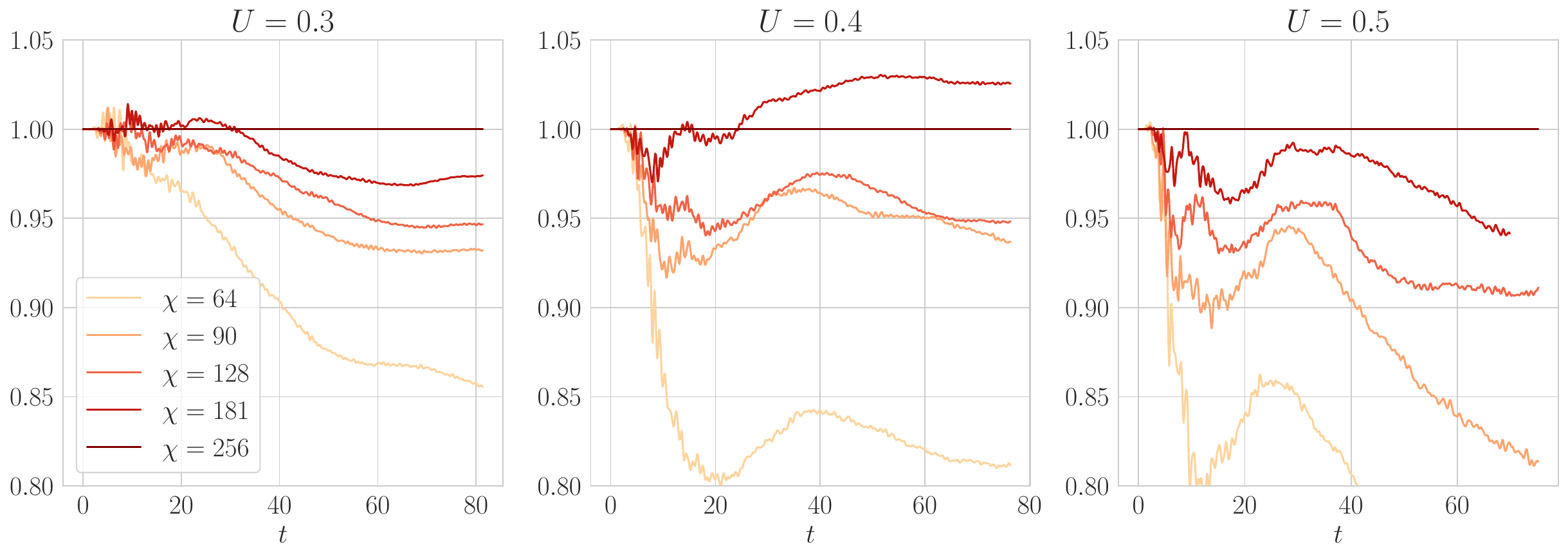}
\caption{\textbf{Top row}: current-current correlator $\langle J(t) J(0)$ of Eq.~\ref{eq:app:DtJJ} computed with DMT for $U = 0.3$ (left), $U = 0.4$ (center), $U = 0.5$ (right).
After a fast initial oscillation, the correlator displays a slow decay well approximated by a two-exponential fit (blue line; Eq.~\ref{eq:app:phenom-two-exp}).
\textbf{Bottom row}: 
convergence in bond dimension.
All curves are at Trotter step $dt = 0.125$}
\label{fig:app:current-decay}
\end{figure*}

\subsection{DMT}

In TEBD
\cite{vidalEfficientClassicalSimulation2003,vidalEfficientSimulationOnedimensional2004,zwolakMixedstateDynamicsOnedimensional2004}
one truncates an MPO with a single SVD, resulting in a local approximation that is optimal with respect to the Frobenius norm.
But the Frobenius norm is blind to the fact that some operators---especially local operators like energy density---are more important than others.

Density matrix truncation \cite{whiteQuantumDynamicsThermalizing2017} replaces the SVD truncation with a truncation that exactly preserves operators with support up to some preservation diameter $l_{\text{pres}}$,
and truncates longer operators via SVD;
it has been successfully applied to thermalizing \cite{} and integrable \cite{} systems.

We implement DMT as modified in \onlinecite{thomasComparingNumericalMethods2023} for Heisenberg dynamics;
for simplicity of implementation, we take a preservation diameter $L_{\text{pres}} = 3$.
We use a second-order boustrophedon (sweeping, DMRG-like) Trotter decomposition,
rather than the usual brickwork Trotter decomposition;
this seems to give better convergence in Trotter step.

\subsection{Current decay}

\subsubsection{Diffusion coefficients and the current-current correlator}
In the main text we extract the diffusion coefficient from the variance of the energy density correlator.
That correlator is
\begin{align}
    C^{\varepsilon\varepsilon}(x,t) = \expct{\varepsilon_x(t)\varepsilon_x(L/2)}\;.
\end{align}
To extract a time-dependent diffusion coefficient we first compute the mean squared displacement
\begin{align}
    V(t) = \frac 1 \nu \sum_x x^2 C^{\varepsilon\varepsilon}(x,t) - \left(\sum_x x C(x,t)\right)^2
\end{align}
where $\nu$ is a normalization
\begin{align}\label{eq:app:C-norm}
    \nu = \sum_{x} C(x,t) = \sum_x C(x,0) = \expct{\varepsilon_{L/2}^2}\;.
\end{align}
The time-dependent diffusion coefficient is
\begin{align}
    D(t) = \frac 1 2 \frac{d}{dt} V(t)\;;
\end{align}
we estimate this via a numerical derivative.
We then estimate the physical diffusion coefficient
\begin{align}
    D = \lim_{t \to \infty} D(t)
\end{align}
by fitting $D(t)$ to the functional form
\begin{align}\label{eq:app:Dt-functional-form}
   D(t) = D - B e^{-t/\tau} \;, \quad t > t_0
\end{align}
after some initial time $t_0$.

It is useful to check the functional form \eqref{eq:app:Dt-functional-form} by computing directly computing the derivative $\frac d {dt} D(t)$.
We can write $\frac d {dt} D(t)$ as a correlator by repeatedly applying the conservation law $\partial_t \varepsilon_x = j_{x-1} - j_x$, summation by parts, time translation invariance, and spatial translation invariance to the correlator $C(x,t)$; the result is
\begin{align}\label{eq:app:DtJJ}
    \frac d {dt} D(t) = \frac 1 \nu \langle J(t) J(0) \rangle
\end{align}
where
\begin{align}
    J(t) = \sum_x j_x(t)\;,
\end{align}
$j_x(t)$ the local energy current operator,
and $\nu$ is the same normalization \eqref{eq:app:C-norm}.

\subsubsection{Results and convergence}

\begin{figure}
\includegraphics[width=0.45\textwidth]{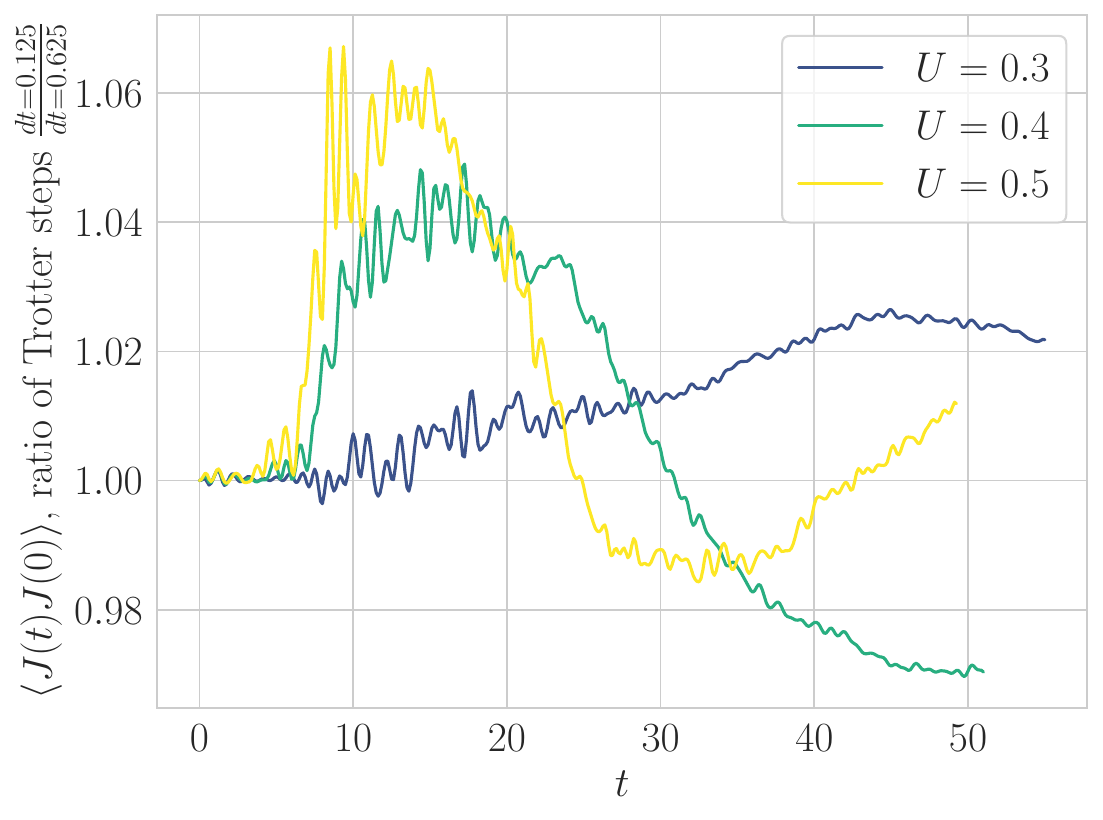}
\caption{}
\caption{Trotter step convergence of the current-current correlator in DMT simulations. Each curve is at bond dimension $\chi = 256$.}
\label{fig:app:current-decay-trotter}
\end{figure}

\begin{figure*}
\includegraphics[width=\textwidth]{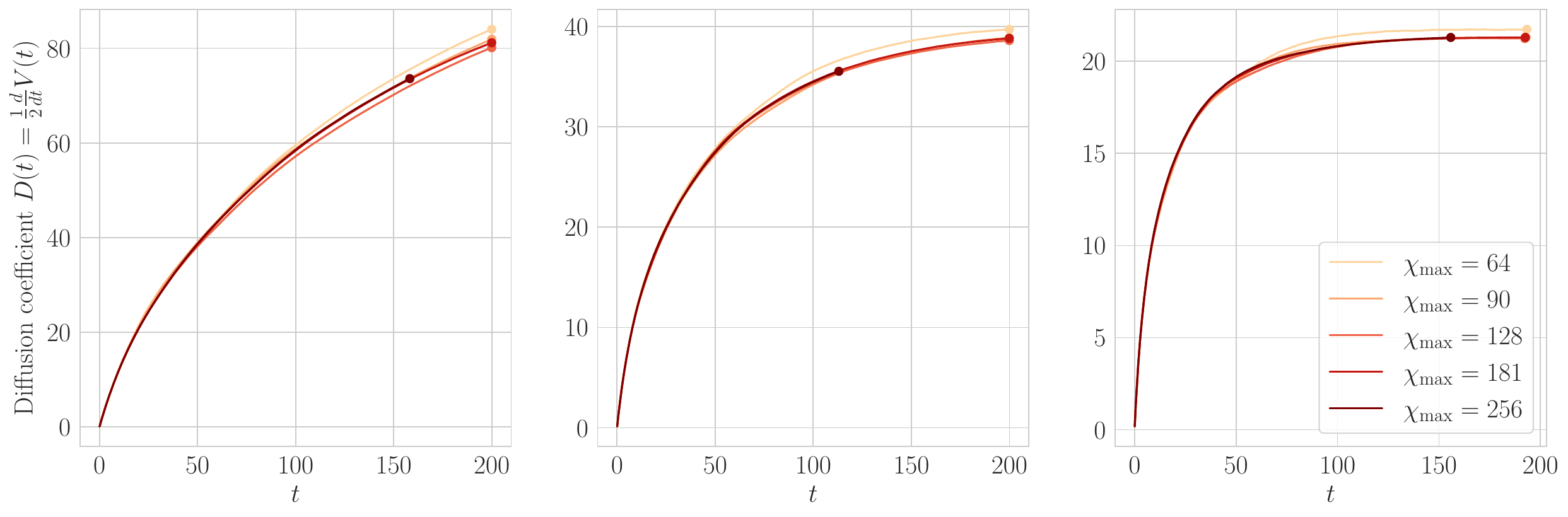}
\includegraphics[width=\textwidth]{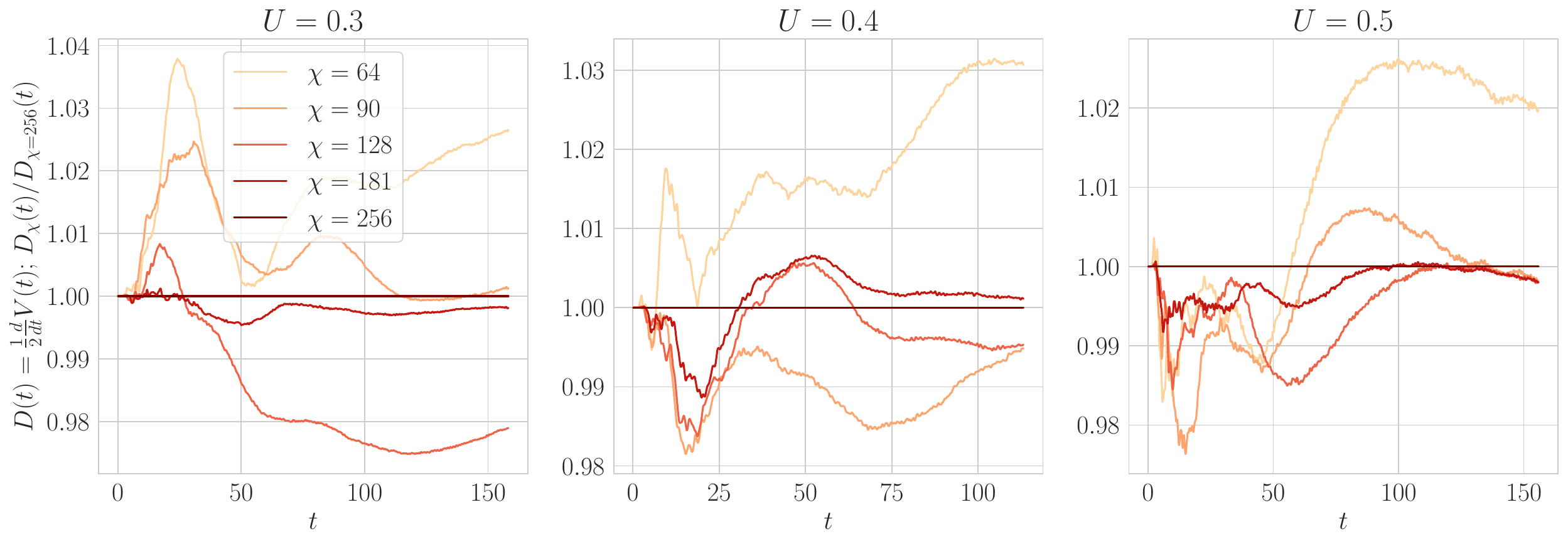}
\caption{\textbf{Top row} time-dependent diffusion coefficients $D(t)$ extracted from DMT simulations of the mean square displacement $V(t)$ for $U = 0.3, 0.4, 0.5$. \textbf{Bottom row:} bond dimension convergence of extracted $D(t)$; in each case $D(t)$ differs between bond dimension $\chi = 64$ and bond dimension $\chi = 256$ by less than $5\%$.
DMT simulations use Trotter step $dt = 0.125$}
\label{fig:app:msd}
\end{figure*}

Fig.~\ref{fig:app:current-decay} shows the current-current correlator $\expct{J(t)J(0)}$ as a function of time for $U = 0.3, 0.4, 0.5$.
In each case the correlator shows fast early oscillations, followed by a slow decay.
The fast oscillations result from the definition of the current (see App.~\ref{app:ss:density-current}).
The slow decay is well-approximated by a phenomenological functional form
\begin{align}\label{eq:app:phenom-two-exp}
    \expct{J(t)J(0)} = A e^{-t/\tau_1} + B e^{-t/\tau_2}\;.
\end{align}
We fit to the $t \ge 5$ data to avoid the initial oscillations, and give the resulting time constants $\tau$ in Table \ref{tab:time-constant}.

\begin{table}[h!]
    \centering
    \begin{tabular}{|ccc|}
    \hline
    $U$ & $\tau_1$ &$\tau_2$ \\
    \hline
     0.3 & 15.3 & 133\\
     0.4 & 6.22 & 44.7\\
     0.5 & 4.38 & 21.45\\
     \hline
    \end{tabular}
    \caption{Time constants for the phenomenological two-exponential fit (Eq.~\ref{eq:app:phenom-two-exp}) to the current-current correlation function \eqref{eq:app:DtJJ} plotted in Fig.~\ref{fig:app:current-decay}}
    \label{tab:time-constant}
\end{table}

Fig.~\ref{fig:app:current-decay} bottom row shows convergence of the DMT current-current correlator in bond dimension for $U = 0.3, 0.4, 0.5$. We plot
\begin{align}
    \frac
    {\expct{J(t) J(0)}[\chi]}
    {\expct{J(t) J(0)}[\chi = 256]}\;.
\end{align}
In each case we find that $C^{JJ}_{\chi = 256}(t)$ is within 10\% of $C^{JJ}_{\chi = 512}$.

This $10\%$ difference understates convergence error, because $C^{JJ}_\chi(t)$ trends upward as the bond dimension $\chi$ increases.
(The trend is unambiguous for $U = 0.3, 0.5$; for $U = 0.4$ it is less clear, but arguably still present.)
It appears that DMT systematically underestimates $C^{JJ}_\chi(t)$.
We believe this underestimate results from our choice of preservation diameter
We use DMT with preservation diameter $3$, meaning it exactly preserves only those operators with support on up to 3 sites,
but the energy current is a 4-site operator.
We believe that simulations with preservation diameter $\ge 4$ would converge more quickly.

Fig.~\ref{fig:app:current-decay-trotter} shows convergence of the DMT current-current correlator in Trotter step $dt$.
We plot
\begin{align}
    \frac
    {\expct{J(t) J(0)}[dt = 0.125]}
    {\expct{J(t) J(0)}[dt = 0.0625]}
\end{align}
for $U = 0.3, 0.4, 0.5$; in each case Trotter step $dt = 0.125$ is within $10\%$ of Trotter step $dt = 0.0625$.

\subsubsection{Definitions of energy density and energy current}\label{app:ss:density-current}

\begin{figure}
\includegraphics[width=0.45\textwidth]{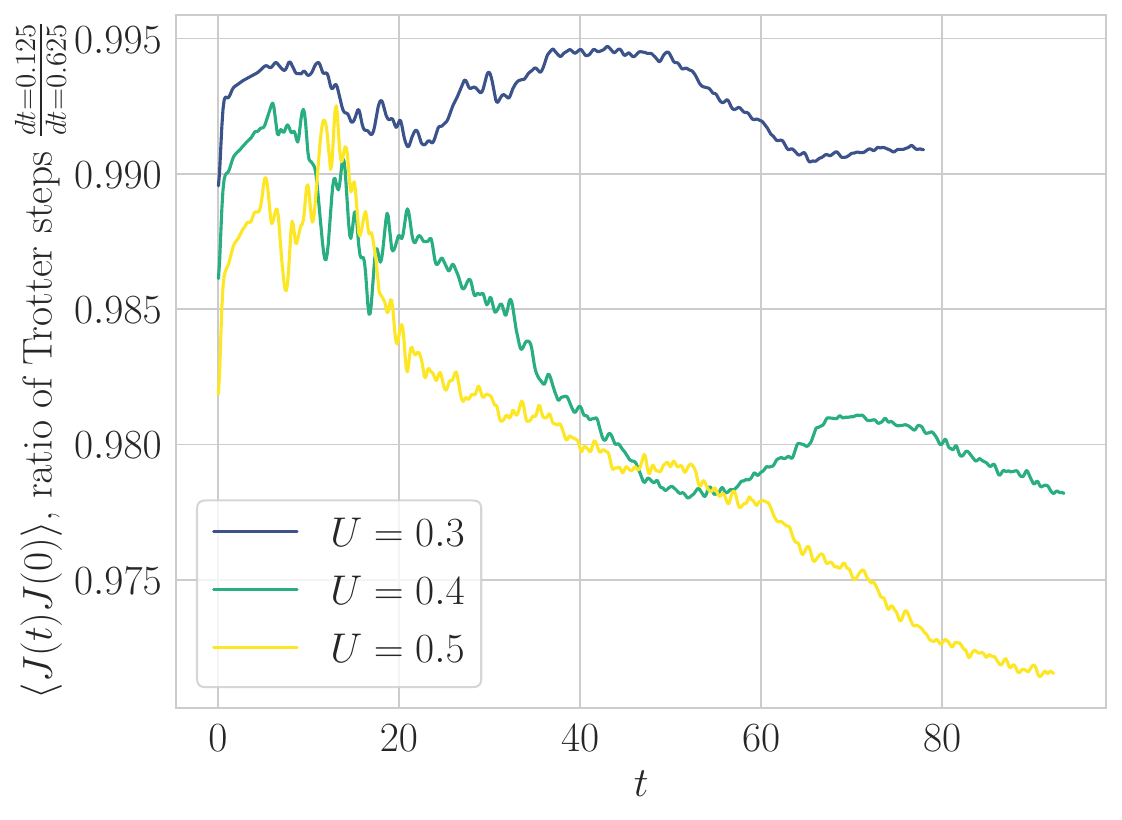}
\caption{MSD Trotter}
\label{fig:app:msd-convergence-trotter}
\end{figure}

In this appendix we have used definitions of energy density and energy density current that are natural in fermion language, rather than in spin language, because the requisite analytical calculations (especially of the current itself) were more convenient in fermion language.
In Majorana language the energy density is 
\begin{align}\label{eq:app:energy-density-majorana}
\varepsilon^{(M)}_\xi &= i \eta_\xi \eta_{\xi+1} - U\eta_{\xi-1}\eta_\xi\eta_{\xi+1}\eta_{\xi+2}\;.
\end{align}
$\varepsilon^{(M)}_\xi$ is symmetric under reflection about the bond $(\xi,\xi+1)$.
The current of this energy density $\varepsilon^{(M)}$ can be written
\begin{align}
\begin{split}
    j^{(M)}_\xi &= -2i\Big[P_\xi - iU (A_{\xi-2} + A_{\xi-1} + B_{\xi-1} + B_\xi) \\
    &\qquad\qquad + U^2 ( D_{\xi-1} - C_{\xi-3} - C_{\xi-2} - C_{\xi-1} ) \Big] \\
    P_\xi &= \eta_\xi \eta_{\xi+2} \\
    A_\xi &= \eta_\xi \eta_{\xi+1} \eta_{\xi+2} \eta_{\xi+4} \\
    B_\xi &= \eta_\xi \eta_{\xi+2} \eta_{\xi+3} \eta_{\xi+4} \\
    C_\xi &= \eta_\xi \eta_{\xi+1} \eta_{\xi+2} \eta_{\xi+4} \eta_{\xi+5} \eta_{\xi + 6} \\
    D_\xi &= \eta_\xi \eta_{\xi+4}\;
\end{split}
\end{align}
where we label the Majorana sites $\xi = 2, \dots, 2L$.

In preparation for Jordan-Wigner transformation it is helpful to group sites:
one grouped site $x$ corresponds to two Majorana sites $(2x,2x+1)$.
The energy density \eqref{eq:app:energy-density-majorana} is then
\begin{align}\label{eq:app:energy-density-majorana-grouped}
\begin{split}
\varepsilon_x
&= \varepsilon^{(M)}_{2x} + \varepsilon^{(M)}_{2x+1} \\
&= \quad i \eta_{2x} \eta_{{2x}+1} - U\eta_{{2x}-1}\eta_{2x}\eta_{{2x}+1}\eta_{{2x}+2} \\
&\quad + i \eta_{2x+1} \eta_{2x+2} - U\eta_{2x}\eta_{2x+1}\eta_{2x+2}\eta_{2x+3}\;;
\end{split}
\end{align}
the current of this energy density can be written
\begin{align}
    j_x = j^{(M)}_{2x}\;.
\end{align}
Note that the total currents are not the same:
\begin{subequations}
\begin{align}
    J &:= \sum_x j_x = \sum_x j^{(M)}_{2x} \label{eq:app:total-current-grouped}\\
    J^{(M)} &:= \sum_\xi j^{(M)}_\xi = \sum_x [ j^{(M)}_{2x} + j^{(M)}_{2x+1} ] \\
    &\ne J \notag\;.
\end{align}
\end{subequations}

The difference between $J$ and $J^{(M)}$ explains the early-time oscillations of the energy current in Fig.~\ref{fig:app:current-decay}.
Take $U = 0$, for simplicity.
In that case one can check
\begin{align}\label{eq:app:total-JM-dt}
    d_t J^{(M)} = 0\;.
\end{align}
If we decompose
\begin{align}
    J^{(M)} = J + J'\;,
\end{align}
$J$ the total current of \eqref{eq:app:total-current-grouped} consisting only of even terms and
\begin{align}
    J' = \sum_x j^{(M)}_{2x+1}
\end{align}
consisting of odd terms, then $d_t J^{(M)} = 0$ implies
\begin{align}
  d_t J = - d_t J'  
\end{align}
leading to oscillations.

This is a lattice-scale phenomenon.
If $U \ne 0$, for any but the shortest times $\varepsilon^{(M)}_{2x} \approx \varepsilon^{(M)}_{2x+1}$ and $j^{(M)}_{2x} \approx j^{(M)}_{2x+1}$, so the decay of $J$ broadly matches that of $J^{(M)}$.

\subsection{Mean square displacement in DMT simulations}

The DMT diffusion coefficients in the main text come from the mean square displacement $V(t)$, analyzed in the same way asthe FDAOE data.
Fig.~\ref{fig:app:msd} top shows the diffusion coefficient extracted from the mean square displacement for $U = 0.3, 0.4, 0.5$;
Fig.~\ref{fig:app:msd} bottom shows convergence in bond dimension, and Fig.~\ref{fig:app:msd-convergence-trotter} shows convergence in Trotter step.
Bond dimension convergence error is $\lesssim 3\%$ for $\chi = 128$ compared to $\chi = 256$
(with better convergence for larger $U$).
Trotter error is also $\lesssim 3\%$ for times $t \lesssim 90$, but growing.

\end{document}